\documentclass[twocolumn,nolinenumbers]{aastex631}

\usepackage{amsmath}
\usepackage{cleveref}

\newcommand\tdisc{t_\mathrm{disk}}
\newcommand\tauc{\tau_\mathrm{c}}
\newcommand\Prot{P_\mathrm{rot}}
\newcommand\Lx{L_\mathrm{X}}

\newcommand\Msun{M_{\mathrm{\odot}}}
\newcommand\Mstar{M_{\mathrm{\star}}}

\newcommand\Mdotwind{\dot{M}_{\mathrm{w}}}
\newcommand\Mdotacc{\dot{M}_{\mathrm{acc}}}
\newcommand\ergs{\mathrm{erg\,s^{-1}}}
\newcommand\tclear{t_{\mathrm{c}}}
\newcommand\tnu{t_\nu}
\newcommand\Rg{R_\mathrm{g}}

\newcommand\Myr{\mathrm{Myr}}
\newcommand\Gyr{\mathrm{Gyr}}
\newcommand\yr{\mathrm{yr}}
\newcommand\days{\mathrm{d}}
\newcommand\au{\mathrm{au}}
\newcommand\pc{\mathrm{pc}}

\defcitealias{Picogna+2019}{P19}
\defcitealias{Garraffo+2018}{G18}
\defcitealias{Garraffo+2015}{G15}
\defcitealias{Garraffo+2016}{G16}

\received{May 26, 2023}
\revised{September 19, 2023}
\accepted{November 05, 2023}

\shorttitle{Linking circumstellar disk lifetimes to the rotational spin-down of low-mass stars}
\shortauthors{Monsch et al.}
\graphicspath{{./}{plots/}}
\begin{document}

\title{Linking circumstellar disk lifetimes to the rotational evolution of low-mass stars}

\correspondingauthor{Kristina Monsch}
\email{kristina.monsch@cfa.harvard.edu}

\author[0000-0002-5688-6790]{K. Monsch}
\affiliation{Center for Astrophysics $\vert$ Harvard \& Smithsonian, 60 Garden Street, Cambridge, MA 02138, USA}

\author[0000-0002-0210-2276]{J. J. Drake}
\affiliation{Center for Astrophysics $\vert$ Harvard \& Smithsonian, 60 Garden Street, Cambridge, MA 02138, USA}
\affiliation{Lockheed Martin, 3251 Hanover St, Palo Alto, CA 94304}

\author[0000-0002-8791-6286]{C. Garraffo}
\affiliation{Center for Astrophysics $\vert$ Harvard \& Smithsonian, 60 Garden Street, Cambridge, MA 02138, USA}

\author[0000-0003-3754-1639]{G. Picogna}
\affiliation{Universit\"ats-Sternwarte, Fakult\"at f\"ur Physik, Ludwig-Maximilians-Universit\"at M\"unchen, Scheinerstr.~1, 81679 M\"unchen, Germany}

\author[0000-0002-5688-6790]{B. Ercolano}
\affiliation{Universit\"ats-Sternwarte, Fakult\"at f\"ur Physik, Ludwig-Maximilians-Universit\"at M\"unchen, Scheinerstr.~1, 81679 M\"unchen, Germany}
\affiliation{Exzellenzcluster `Origins', Boltzmannstr. 2, 85748 Garching, Germany}

\begin{abstract}

The high-energy radiation emitted by young stars can have a strong influence on their rotational evolution at later stages. This is because internal photoevaporation is one of the major drivers of the dispersal of circumstellar disks, which surround all newly born low-mass stars during the first few million years of their evolution.
Employing an internal EUV/X-ray photoevaporation model, we have derived a simple recipe for calculating realistic inner disk lifetimes of protoplanetary disks. This prescription was implemented into a magnetic morphology-driven rotational evolution model and is used to investigate the impact of disk-locking on the spin evolution of low-mass stars. 
We find that the length of the disk-locking phase has a profound impact on the subsequent rotational evolution of a young star, and the implementation of realistic disk lifetimes leads to an improved agreement of model outcomes with observed rotation period distributions for open clusters of various ages. 
However, for both young star-forming regions tested in our model, the strong bimodality in rotation periods that is observed in h\,Per could not be recovered. h\,Per is only successfully recovered, if the model is started from a double-peaked distribution with an initial disk fraction of $65\,\%$. However, at an age of only $\sim 1\,\Myr$, such a low disk fraction can only be achieved if an additional disk dispersal process, such as external photoevaporation, is invoked. These results therefore highlight the importance of including realistic disk dispersal mechanisms in rotational evolution models of young stars.

\end{abstract}




\section{Introduction} 
\label{sec:intro}

Planets form in circumstellar disks that surround their host stars during the first few million years of their evolution. Since their first detection in the mid to late 1980s \citep[e.g.][]{SmithTerrile1984, BackmanGillett1987, Strom+1989}, disk properties have been well established. Observational studies using different tracers, such as near-infrared (NIR) or H$\alpha$-excess emission, have shown that the majority of circumstellar disks dissipate after about $\sim 1$--$3\,\Myr$, with only few disks surviving much longer than $10\,\Myr$ \citep[e.g.][]{Haisch+2001, Mamajek+2009, Fedele+2010, Ribas+2014, Ribas+2015}.
More recently, this classic picture seems to have been challenged by studies recalculating disk fractions for a number of star-forming regions, finding that median disk lifetimes can be extended by a factor of 2--3, when correcting for selection effects \citep[e.g.][]{Michel+2021, Pfalzner+2022}. Further, theoretical disk longevity estimates appear to be strongly affected by the choice of pre-main sequence (PMS) evolutionary models \citep[][]{Richert+2018}.

This early phase, in which the central star is thought to be still magnetically coupled to its surrounding disk, is not only important to nascent planets by providing the necessary density of planetary building blocks, but also by preventing the central stars from spinning up through accretion and contraction toward the main sequence in a process known as `disk-locking' \citep[][]{Koenigl1991, CollierCameronCampbell1993, Edwards+1993, Shu+1994b, OstrikerShu1995}. 

The rotation history of a T~Tauri star can carry through to its main sequence phase for hundreds of Myr to a Gyr or so, until rotation has converged to the mass-dependent `Skumanich branch' \citep[$\Prot \propto t^{0.5}$,][]{Skumanich1972}, suggesting that for ages $\gtrsim 1\,\mathrm{Gyr}$, the memory of the early rotational properties is erased. 
However, since rotation powers the magnetic activity responsible for UV-, X-ray and energetic particle radiation, the early phases of rotation will be crucial for the integrated radiation dose of planets and their capacity to host and retain an atmosphere. Thus, understanding the physical processes that are present during this early phase of stellar evolution has tremendous importance for studies of planetary formation \citep[e.g.][]{Emsenhuber+2023}, their potential habitability \citep[e.g.][]{Johnstone+2015, Gallet+2017} and the ability to detect such habitable systems \citep[e.g.][]{Newton+2016}.

\subsection{Magnetic field morphology of young stars}
\label{sec:intro_magneticmorphology}

Zeeman-Doppler-Imaging (ZDI) observations have shown that young and active stars likely store a significant fraction of their magnetic flux in higher-order multipole components of their magnetic fields, meaning that they are magnetically more complex \citep[e.g.][]{Folsom+2016}. 
The importance of magnetic field morphology (which is sometimes referred to as topology) on stellar evolution has only recently been a subject of detailed research. \citet{Reville+2015} provided torque formulations for different magnetic field morphologies, finding that the more complex the magnetic field the smaller the torque is.
Almost at the same time, the study by \citet[][henceforth \citetalias{Garraffo+2015}]{Garraffo+2015} found high-order surface fields had greatly reduced mass loss and torques compared with low-order fields. They suggested that multipolar magnetic field configurations can provide the physical basis for the so-called Metastable Dynamo Model (MDM) proposed by \citet{Brown2014} that aimed at explaining the bimodality in rotation periods in young open clusters due to a sudden transition from a weak to strong coupling of the magnetic field with the stellar wind. 

The later study by \citet[][henceforth \citetalias{Garraffo+2016}]{Garraffo+2016} provided a physical foundation for the MDM by including higher-order magnetic field complexities in stellar rotational evolution models. They found that this can lead to a reduction of up to three orders of magnitude for the angular momentum loss (AML) rates. The reason for this is twofold. Firstly, with higher magnetic complexity the area coverage of coronal holes and open magnetic field from which the fast and less dense winds originate decreases. Secondly, with higher complexity the magnetic field strength drops off more rapidly with radial distance and the magnetic lever arm acting on the slow wind can be greatly decreased.
This finding is nevertheless still consistent with studies showing that the dominant component in AML is the dipolar magnetic field \citep[e.g. \citetalias{Garraffo+2016},][]{FinleyMatt2018, See+2019}, since for given magnetic flux, higher complexities imply a lower fraction of magnetic flux in the dipolar component, which remains, however, the most efficient for AML.

Initially fast-rotating stars tend to show higher complexities in their large scale magnetic field structure \citep[e.g.][]{Folsom+2016}, so consequently they lose angular momentum less efficiently due to their smaller number of open magnetic field lines and the shorter range of the magnetic lever arm of higher order fields \citepalias{Garraffo+2015, Garraffo+2016}.
This leads to a more extended period of time in which the rotational velocities of these stars decrease, while their magnetic complexity is being slowly eroded. Due to the steep dependence of AML with magnetic complexity \citepalias{Garraffo+2015}, this ultimately results in a sharp transition from fast to slow rotation before these stars converge towards the Skumanich-branch that can be well described by AML due to a bipolar magnetic field configuration. 
In contrast, initially slowly rotating stars generally show less complex magnetic configurations, therefore they can lose mass and angular momentum more efficiently. Such stars quickly converge towards the slow-rotating branch without experiencing a rapid spin-down phase.

\subsection{Disk-locking}
\label{sec:intro_disklocking}

`Disk-locking' describes a magnetic coupling between the host star's magnetosphere and the surrounding disk, which is hypothesized to enable the system to transfer and lose angular momentum, and therefore effectively brake the star's spin-up resulting from accretion from the disk and contraction in the PMS phase \citep[see however][]{Matt+2010, Matt+2012}.
While observations of young clusters with ages near the zero-age main sequence (ZAMS) show that low-mass stars exhibit a wide range of rotation periods, their rotational speeds stay almost constant during the interaction phase with their circumstellar disks \citep[e.g.][but see also \citealt{Gallet+2019}]{GalletBouvier2013, GalletBouvier2015}. Once these are dispersed, the range of rotation periods narrows down for older main sequence clusters \citep[e.g.,][]{StaufferHartmann1987, Prosser+1995, Adams+1998}. 
Further, it has been shown that the bimodality in rotation periods that is observed in some of these young star-forming regions, with disk-bearing stars showing--on average--slower rotation than disk-less stars, is in good agreement with the disk-locking mechanism \citep[e.g.][]{Stassun+2001, Herbst+2002, Rebull2001, Rebull+2004}.

So far, only few studies have investigated in detail the impact of the disk-locking phase on the subsequent rotational evolution of a young star. 
For instance, \citet{Vasconcelos+2015} and \citet{Vasconcelos+2022} studied the initial conditions of a young star-forming region needed in order to reproduce the rotational period distribution of older clusters under the assumption of disk-locking operating. They found that assuming a bimodal distribution of periods, with disk-less stars rotating on average faster than disk-bearing ones, as well as including realistic scaling relations for various other stellar properties can reproduce the bimodal period distribution observed in the $\sim 13\,\Myr$ old h\,Per cluster. 

On the other hand, \citet{Roquette+2021} investigate how the local FUV-environment around massive stars can alter circumstellar disk lifetimes and thus affect the rotational history of low-mass stars. These authors find that the radiation-environment leaves an imprint in the period-mass distributions until past the early PMS phase, even up to the main sequence and that the excess of fast rotators in h\,Per can be explained by a high FUV environment during the cluster’s early PMS evolution. These authors do not include the effect of internal circumstellar disk photoevaporation by high-energy photons on the spin-evolution process of their host stars.
However, compared to their main sequence counterparts, PMS stars are known to show highly elevated levels of X-ray emission \citep{Feigelson+2007_PP5}. There is growing evidence that X-rays from the central star may be the major driver of circumstellar disk evolution \citep[e.g.][]{Ercolano+2023}, as they heat their upper layers, which ultimately leads to the formation of centrifugally launched, thermally-driven winds that disperse the disk from the inside out \citep[cf.][for recent reviews]{ErcolanoPascucci2017, ErcolanoPicogna2022}.

In this paper, we explore the combined effect of disk-locking and stellar magnetic morphology on the subsequent rotational evolution of low-mass stars. 
We include self-consistent calculations of inner circumstellar disk lifetimes based on stellar X-ray emission into an extension of the stellar spin-evolution model of \citet[][henceforth \citetalias{Garraffo+2018}]{Garraffo+2018} that takes higher multipole components of the stellar magnetic fields into account.
We apply this model to evolve the rotational evolution of partly-convective, low-mass stars ($0.4$--$1.1\,\Msun$) all the way from the early PMS phase to roughly solar age.
In \S\,\ref{sec:methods}, we describe the techniques and numerical model employed in our study. The results and discussion are presented in \S\,\ref{sec:results} and \S\,\ref{sec:discussion}, respectively. 
The conclusions follow in \S\,\ref{sec:conclusions}.
\section{Methods} 
\label{sec:methods}

\subsection{Calculation of inner disk lifetimes}
\label{sec:methods_tdisc}

The inner disk lifetimes\footnote{In this work, we use the terms `inner disk lifetime' and `disk-locking time' interchangeably.} are calculated based on the internal EUV+X-ray photoevaporation (XPE) model presented by \citet{Picogna+2019, Picogna+2021} and \citet{Ercolano+2021}. Within this framework, circumstellar disks disperse via a combination of viscous accretion and thermal disk winds that are launched due to internal photoevaporation driven by the host star. 

While viscous accretion is expected to dominate the disk's surface density evolution for most of its lifetime, once the accretion rate becomes comparable to the mass loss rate due to the photoevaporative wind, internal photoevaporation will be the main driver of its dispersal. At this point, 
an annular cavity opens close to the gravitational radius, $R_\mathrm{g} = G\Mstar/c_\mathrm{s}^2$, where $G$ is the gravitational constant, $\Mstar$ is the stellar mass and $c_\mathrm{s}$ is the thermal sound speed of the gas. From this radius on outward, vigorous disk winds are established that lead to the fast inside-out dispersal of the circumstellar disk within a few $10^5$ years, which is in good agreement with the so-called `two-timescale behavior' that is typically observed in planet-forming disk statistics \citep[e.g.][]{Clarke+2001, Ercolano+2011, Koepferl+2013}.

While the heating of the disk's surface due to the irradiation by the central star is primarily driven by FUV \citep[$h\nu \approx 6$--$13.6\,$eV, e.g.][]{Gorti+2009a, Gorti+2009b}, EUV \citep[$h\nu \approx 13.6$--$100\,$eV, e.g.][]{Font+2004, Alexander+2006a, Alexander+2006b} and X-ray photons \citep[$h\nu \approx 0.1$--$10\,$keV, e.g.][]{Owen+2010, Owen+2011, Owen+2012, Picogna+2019, Picogna+2021}, recent studies have shown that the internal photoevaporation rates are predominantly controlled by the luminosities in the soft X-ray band \citep[$h\nu \approx 0.1$--$1\,$keV, cf.][see, however, \citealt{WangGoodman2017} and \citealt{Nakatani+2018}]{Ercolano+2008, Ercolano+2009a, Ercolano+2021}. 

The lifetime of the circumstellar disk inside the gravitational radius can then be expressed as the sum of the clearing timescale $\tclear$ \citep[i.e. the time at which photoevaporation starts to clear the circumstellar disk and $\Mdotwind \approx \dot{M}_\mathrm{acc}$, cf.][]{Clarke+2001, Ruden2004} and the viscous timescale, $\tnu$, at the gap-opening radius, $R_\mathrm{g}$:
\begin{equation}
\label{eq:tdisc}
    \tdisc = \tclear + \tnu (\Rg),
\end{equation}
where
\begin{equation}
\label{eq:tclear}
    \tclear = \frac{M_\mathrm{d,0}}{2\dot{M}_\mathrm{acc,0}^{1/3} \dot{M}_\mathrm{w}^{2/3}}, 
\end{equation}
and
\begin{equation}
\label{eq:tnu}
    \tnu = \frac{R^2}{3\nu} = \frac{M_\mathrm{d,0}}{2\dot{M}_\mathrm{acc,0}}, 
\end{equation}
Here, $M_\mathrm{d,0}$ describes the initial disk mass, $\dot{M}_\mathrm{acc,0}$ the initial accretion rate onto the star and $\dot{M}_\mathrm{w}$ the mass-loss rate due to the photoevaporative winds. 
The full derivation of Eqs.~\ref{eq:tclear} and \ref{eq:tnu} is demonstrated in detail in Appendix~\ref{sec:appendix_tdisc} and results in the fact that the disk lifetime of a given star can be solely calculated from its observed stellar mass and X-ray luminosity when assuming empirical scaling laws. An online-tool\footnote{\url{https://github.com/kristina-monsch/disk_lifetime_calculator}} has been made available that allows users to quickly and easily estimate inner disk lifetimes for arbitrary stellar mass/X-ray luminosity combinations.

We emphasize that Eq.~\ref{eq:tdisc} only describes the \textit{inner} disk lifetime, i.e. the time at which photoevaporation starts to disperse the circumstellar disk by opening an annular cavity at $\Rg$. We assume this to be the only relevant timescale for stellar rotation studies, as photoevaporation fully decouples the inner from the outer disk \citep[somewhere between $\sim1$--$10\,\au$, e.g.][]{Liffman2003, Owen+2012, Picogna+2019}, so that  mass can no longer be transferred from the outer to the inner disk.

\subsection{Magnetic morphology-driven rotational evolution model}
\label{sec:methods_rot}

Our rotational evolution model closely follows the one presented by \citet[][henceforth \citetalias{Garraffo+2018}]{Garraffo+2018}. 
It is based on detailed 3D magneto-hydrodynamical (MHD) simulations presented in \citetalias{Garraffo+2015}, for which a simple analytical recipe had been derived in \citetalias{Garraffo+2016}.
This model accounts for the complexity of stellar surface magnetic fields and the finding that the efficiency of magnetic braking depends on the topology of the stellar magnetic field and not just its strength.
They find that for a given magnetic moment, the mass loss and AML rates are independent of the way in which magnetic flux is distributed across the stellar surface for a given magnetic complexity (i.e. the degree of their axissymmetry) and that the latter is therefore the primary factor affecting AML. 
This model successfully recovers the bimodal distribution of both slow and fast rotating stars that is observed in a variety of young open clusters (OCs) with ages $\gtrsim 100\,\Myr$ \citep[e.g.][]{Meibom+2011}.
We refer the reader to these papers for a detailed description of the underlying physical model, and here only briefly summarize the concept and the most relevant equations. 

The analytic formula that describes the AML is given by
\begin{equation}
    \dot{J}(n)=\dot{J}_\mathrm{dip} Q_\mathrm{J}(n),
\label{eq:dotJ}
\end{equation}
where, $\dot{J}_\mathrm{dip} = J_0 \tauc \Omega^3 $ is the AML rate assuming a dipolar magnetic morphology, $J_0 = 1\times10^{41}\,\mathrm{g\,cm^2}$ is a normalization constant, that is chosen in order to match the current rotation period of the Sun, $\tauc$ is the convective turnover time and $\Omega$ is the stellar rotation rate.

The modulating factor
\begin{equation}
    Q_\mathrm{J}(n) = 4.05 e^{-1.4n} + \frac{n-1}{60 B n} \approx 4.05 e^{-1.4n}
\label{eq:Qn}
\end{equation}
accounts for the complexity of the magnetic fields on the stellar surface, $n$, and the magnetic field strength, $B$.
The right side of Eq.~\ref{eq:Qn} then follows from imposing $n=7$ as the maximum complexity, at which AML rates in simulations reach a plateau \citepalias{Garraffo+2016}. This is consistent with the finding of \citet[][]{Garraffo+2013} who show that the small scale field, i.e. magnetic active regions, do not significantly alter the global structure of the stellar wind.  

Further, the order of magnetic complexity is assumed to be correlated with the Rossby number, $\mathrm{Ro} = \Prot/\tauc$, via

\begin{equation}
    n = \frac{a}{\mathrm{Ro}} + b\mathrm{Ro} + 1,
\label{eq:n}
\end{equation}
with $a=0.02$ and $b=2$. The functional form of Eq.~\ref{eq:n} was chosen such that it reflects the trends seen both in OC and \textit{Kepler} observations from the \textit{K2} mission. 
We note that \citet{Gossage+2021} find $a=0.03$, $b=0.5$ and $J_0 = 3 \times 10^{41}\,\mathrm{g\,cm^2}$ better resemble the trends observed in OC observations when included in stellar evolution models that include the effects of rotation. 

We obtain interpolated stellar properties as a function of age (moment of inertia, stellar radius and convective turnover times) from the \textit{MIST} stellar evolutionary tracks\footnote{Convective turnover times are not by default generated as an output in the \textit{MIST}-tables. The tracks used in this study result from a private communication (Dotter, priv. commun.) and are available for download at \url{https://github.com/cgarraffo/Spin-down-model/tree/master/Mist_data}. The initial properties used to generate the tracks are $[\mathrm{Fe/H}]=0.0$ and $v/v_\mathrm{crit}=0.4$.} \citep{Choi+2016, Dotter+2016} and use these as an input to our model, which begins at $2\,\Myr$ (i.e. the youngest age of the clusters studied in this paper). 
We assume solid-body rotation for the stars in our model and evolve the initial stellar rotation period self-consistently at each subsequent timestep following Eq.~\ref{eq:dotJ} and the relation $\Prot = 2\pi/\Omega$. 
After a given disk-locking time, which is calculated using our photoevaporation model described in Sect.~\ref{sec:methods_tdisc} and during which the stellar rotation period is assumed to stay constant, the stars are free to evolve their rotation period based on the above described model.

Stars close to break-up likely experience a significant reduction of their mass due to the shedding of their outer layers while at the same time expanding radially, which may possibly lead to a drastic spin-down phase that could ultimately protect such a star from breaking entirely apart.
However, the \textit{MIST}-tables used for this study are based on non-rotating MESA models and are thus not able to adequately predict the behavior of stars that are close to break-up. 
If not explicitly stated otherwise, we therefore remove all evolutionary tracks in which the star exceeds its corresponding break-up velocity at any given time-step:
\begin{equation}
    \label{eq:breakup_velocity}
    \Omega_\mathrm{break-up} = \sqrt{\frac{G M_\star}{R_\star^3}},
\end{equation}
where $G$ is the gravitational constant, and $M_\star$ and $R_\star$ are the stellar mass and radius, respectively. We note, however, that this criterion is met in less than $1\,\%$ of our simulations, as it generally only applies to stars with initial rotation periods of $\leq 1\,\mathrm{d}$ at 1--$2\,\Myr$.  
Further, we restrict our analysis to partly-convective stars ($0.4$--$1.1\,\Msun$), for which the dynamo activity is thought to originate in layers at the base of the convective zone \citep[see e.g.][for a review]{Charbonneau2010}.

\section{Results and discussion} 
\label{sec:results}

\subsection{Disk lifetimes}
\label{sec:results_tdisc_calculation}

\begin{figure}
    \centering
    \includegraphics[width=\linewidth]{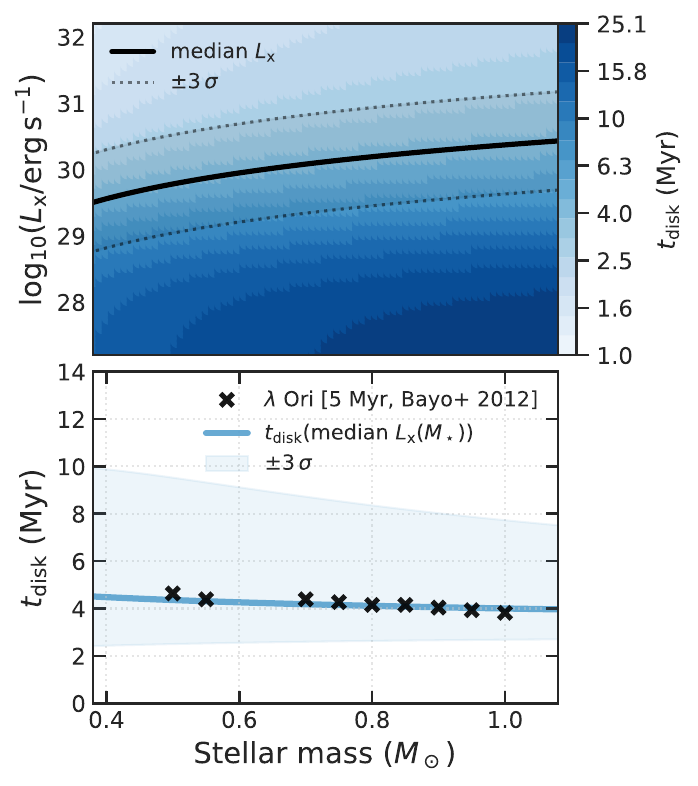}
    \caption{\textit{Top:} Grid of inner disk lifetimes as a function of $\Mstar$ and $\Lx$ where the color-coding reflects the corresponding logarithmic inner disk lifetime $t_\mathrm{disk}(\Mstar, \Lx)$. The black line shows the disk lifetime for the median $\Lx$ for given $\Mstar$ calculated using relation Eq.~\ref{eq:Guedel}. \textit{Bottom:} Inner disk lifetime as a function of $\Mstar$ determined from our model as described in Sect.~\ref{sec:methods_tdisc} (blue line) compared to observed disk fractions from \citet{Bayo+2012} that were converted to disk lifetimes using the method described in \citet{Komaki+2021} and \citet{Picogna+2021}. The shade encompasses the $\pm 3\sigma$-range determined using the XLF for Taurus.}
    \label{fig:disc_life}
\end{figure}

The top panel of Fig.~\ref{fig:disc_life} shows the inner disk lifetimes as a function of stellar mass, $\Mstar$, and X-ray luminosity, $\Lx$, that were calculated using the approach described in Sect.~\ref{sec:methods_tdisc}. 
The black line shows the X-ray luminosity function (XLF) of the Taurus star-forming region \citep[cf. Eq.~\ref{eq:Guedel} and][]{Guedel+2007}, in order to highlight the parts of this diagram which show realistic combinations of $\Lx$ and $\Mstar$ \citep[due to an intrinsic spread of 2--3 orders of magnitude for $\Lx$ for given $\Mstar$, cf.][]{Preibisch+2005, Guedel+2007}.
The color-coding reflects the resulting inner disk lifetimes using a logarithmic scaling.
It can be clearly seen that the inner disk lifetimes strongly decrease with increasing $\Lx$ (for given $\Mstar$) and decreasing $\Mstar$ (for given $\Lx$). However, the dependence on the latter is almost negligible compared to the dependence on $\Lx$, which results in differences in inner disk lifetimes of more than 1 order of magnitude.
This shows that in our model the stellar X-ray luminosity is the primary factor determining the efficiency of XPE, and thus the final dispersal of a circumstellar disk.

The bottom panel of Fig.~\ref{fig:disc_life} essentially shows a collapsed view of the top panel, where the inner disk lifetime is now plotted against the stellar mass. 
The blue line corresponds to the inner disk lifetime assuming the mean $\Lx$ for a given $\Mstar$ determined by the XLF, whereas the light-blue shaded region encompasses the $\pm 3\sigma$ range, determined using the corresponding lower and upper values from the above mentioned XLF.
These are then directly compared to observational data from the $\sim 5\,\Myr$ old $\lambda$\,Ori star-forming region, where the observed disk fractions as a function of $\Mstar$, as determined by \citet{Bayo+2012}, were converted into inner disk lifetimes following the approach suggested by \citet[][see their Sect.~4.1]{Komaki+2021}.

We find that the inner disk lifetimes that are inferred using our model are in good agreement with observational estimates, showing the ability of internal photoevaporation models to reproduce fundamental circumstellar disk statistics.
Further, a modest increase of median disk lifetimes towards lower stellar masses is observed (which would become even more apparent if stars with $M \leq 0.4\,\Mstar$ were to be included in this study), which is consistent with theoretical expectations and observed disk fractions of M-dwarfs \citep[e.g.][]{KennedyKenyon2009, LuhmanMamajek2012, Ribas+2015} and is a direct consequence of lower-mass stars having lower mean $\Lx$.

\subsection{Exploring the individual effect of stellar mass and disk-locking on a star's rotational evolution}

\subsubsection{General rotational evolution}
\label{sec:results_generalevol}

\begin{figure*}
    \centering
    \includegraphics[width=\linewidth]{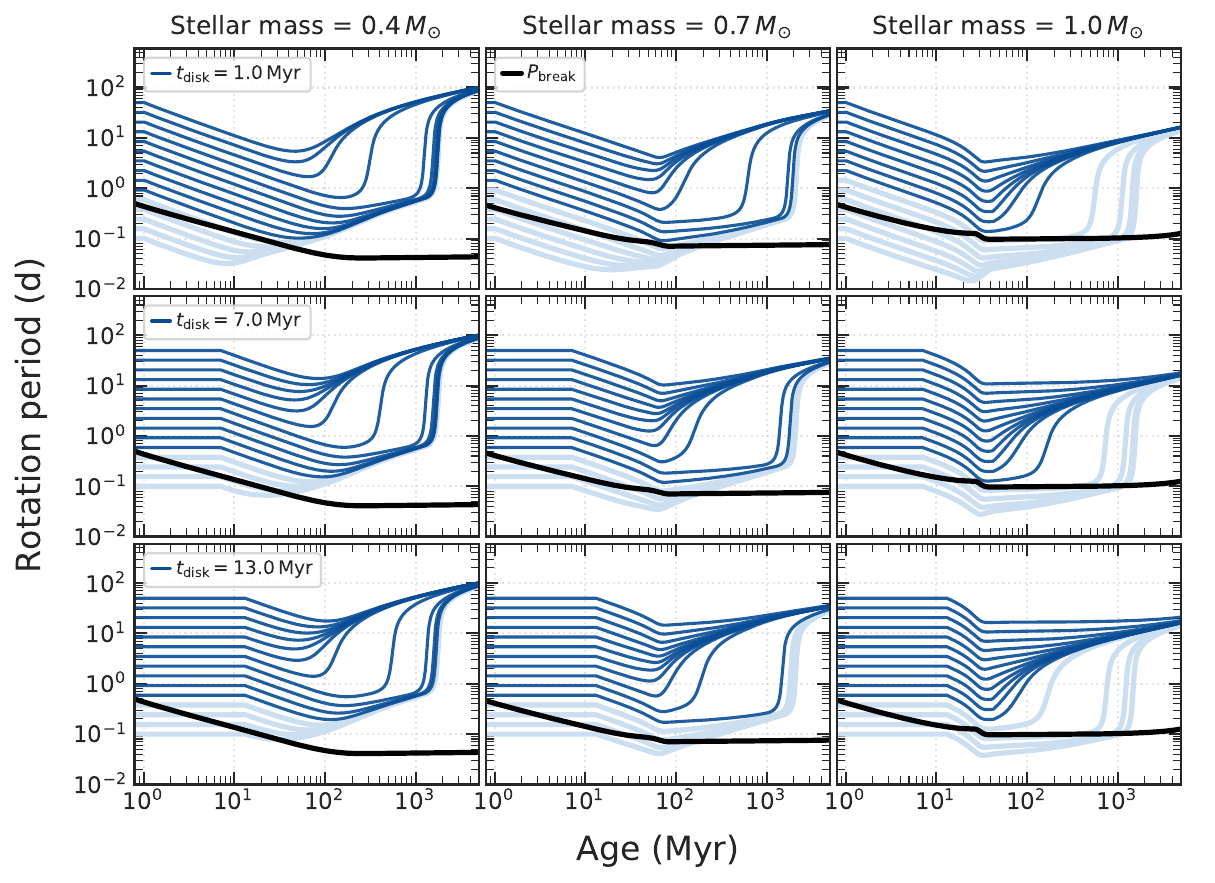}
    \caption{Grid of rotational evolution models as a function of age for different stellar masses and disk-locking times. The columns show the models for $0.4$, $0.7$ and $1\,\Msun$, respectively, while the rows show the models for disk-locking timescales of 1, 7 and $13\,\Myr$. The light-blue lines reflect the evolutionary tracks in which the stars would exceed their corresponding break-up velocity, which is overplotted as black line for each stellar mass.}
    \label{fig:grid}
\end{figure*}

Fig.~\ref{fig:grid} shows a grid of rotational evolution models for stars of $\Mstar = 0.4$, 0.7 and $1\,\Msun$, assuming three different disk-locking timescales of $\tdisc = 1$, 7 and $13\,\Myr$. 
In order to cover a wide range in parameter space, the input rotation periods were sampled uniformly in logarithmic space between $0.1$--$50\,\days$. We note, however, that the observed rotation period distributions of young clusters are significantly more narrow, typically spanning only $\sim 0.5$--$20\,\mathrm{d}$ \citep[see also][]{Gehrig+2023a}. 
We also note that some of the faster rotating evolutionary tracks, colored light-blue, belong to stars whose rotational speeds exceed their corresponding break-up value (denoted by the black line) at some point during their evolution. 
While in reality these tracks are not strictly viable, it is worth keeping them in Fig.~\ref{fig:grid} for a parameter-space investigation of our model. Fig.~\ref{fig:grid} should therefore not be directly compared to observed distributions of open cluster rotation periods. 

For all models, the characteristic evolution of the stellar rotation period, as was already reported in previous studies \citep[e.g.][]{Spada+2011, GalletBouvier2013, GalletBouvier2015, Tu+2015, Amard+2016, Amard+2019, Gondoin2018, Johnstone+2021, Gehrig+2022, Gehrig+2023a, AmardMatt2023}, is observed. 
While the rotation period stays constant throughout the disk-locking phase (which is a common assumption made in many stellar rotation models, see e.g. \citealt{Bouvier+1997, Tinker+2002, Rebull+2004}, but see also \citealt{Gallet+2019}), once the disk is fully dispersed, the stars are free to spin up due to their ongoing contraction and the development of a radiative core, resulting in a decrease of their moment of inertia.
After a few tens of $\Myr$, most stars will have reached the ZAMS, and therefore their contraction phase comes to an end. 
From this point on, their spin-down process is dominated by `magnetic braking' \citep[e.g.][]{WeberDavis1967}, where the magnetic coupling between the star and stellar wind leads to a gradual decrease in rotational velocity with age.

The hitherto almost parallel rotation tracks now begin to diverge into two separate populations---a fast-rotating and a slow-rotating (or Skumanich-) branch.
In our model, this bifurcation is the result of differences in the morphology, and especially the changing complexity of the stellar surface magnetic fields and the resulting varying efficiency of the AML. The numbers of stars residing in either of those branches at a given time is, next to the initial distribution of rotation periods, a function of both their mass and their disk-locking times, which will be addressed in more detail in Sects.~\ref{sec:results_Mstar} and \ref{sec:results_tdisc_model}.


\subsubsection{Stellar mass}
\label{sec:results_Mstar}

For a given disk-locking time (see the rows in Fig.~\ref{fig:grid}), the M-dwarfs in our study ($M_\star \lesssim 0.5\,\Msun$) are more frequently located in the fast-rotating branch for a wide range of initial rotation periods ($\lesssim 10\,$d). Only the slowest rotators remain in the upper, slow-rotating branch.
In contrast, for solar-type stars (i.e. stars with $0.5\,\Msun < M_\star < 1.1\,\Msun$) only the initially fastest rotators ($\Prot \lesssim 3\,$d) evolve towards the fast-rotating branch, while all remaining stars spin down efficiently and converge towards the Skumanich-branch.
As described by \citetalias{Garraffo+2018}, this difference in evolution based on stellar mass can be related to the deeper convective envelopes and longer convective turnover times of low-mass stars, which in turn leads to more complex magnetic field topologies and reduced AML rates compared to high-mass stars which have thinner convective envelopes.

\subsubsection{Disk lifetime}
\label{sec:results_tdisc_model}

Stars lose the majority of their angular momentum during the first few Myr of their evolution, in which they are surrounded by their circumstellar disks. Therefore, stars that are braked for a longer period of time (see the columns in Fig.~\ref{fig:grid}) lose a larger amount of angular momentum during this early PMS phase, resulting in larger rotation periods (i.e. slower rotation) on the ZAMS compared to coeval stars with shorter-lived disks \citep[][]{Eggenberger+2012}. 
In each column of Fig.~\ref{fig:grid}, this can be inferred from the shift of the dip occurring at $\sim 100\,\Myr$ towards earlier ages for shorter disk lifetimes, 
as well as the larger amount of stars ending up in the upper, slow-rotating branch for longer disk-locking times.

Finally, it is worth noting that the bimodality in rotation periods that is produced by our model for ages $\gtrsim 1\,\mathrm{Gyr}$ would only be observable for masses $\lesssim 0.9\,\Msun$ or so, as for higher masses, most of the stars populating the fast-rotating branch would exceed their break-up speed (and thus be removed from our analysis).
The number of stars exceeding break-up is reinforced with shorter disk-locking times, due to these stars having more prolonged spin-up phases resulting from the higher amount of available angular momentum at the end of their ZAMS phase, as well as higher masses, due to their break-up rotation periods decreasing much less steeply compared to lower-mass stars.

\subsection{Constant vs. variable disk-locking times: testing young star-forming regions}
\label{sec:const_vs_var_SFRs}

\begin{figure*}[ht!]
    \centering    
    \includegraphics[width=0.95\linewidth]{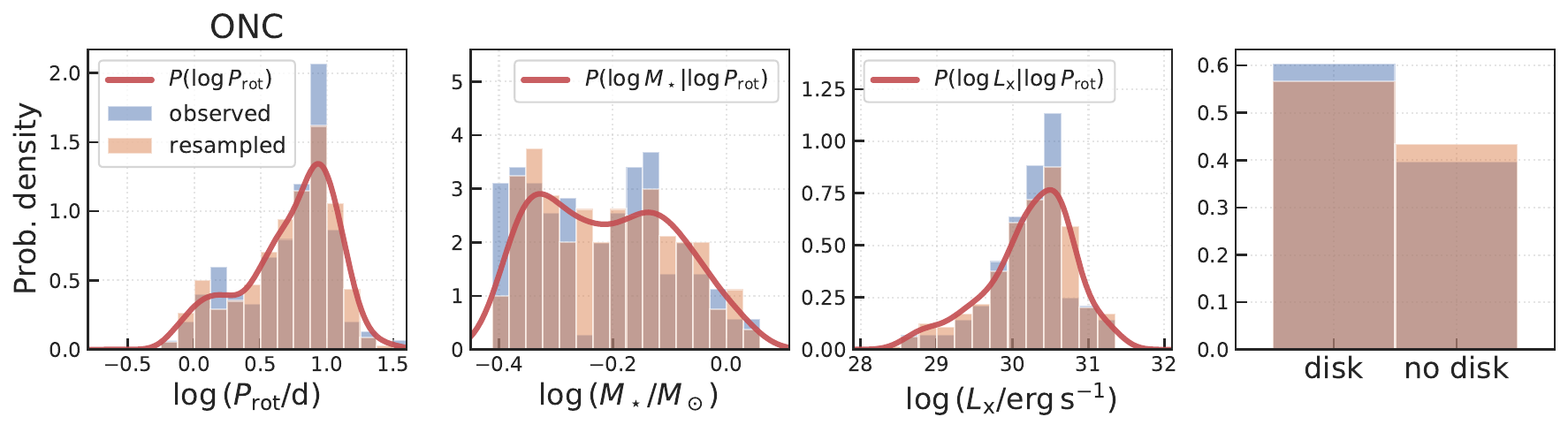}
    \includegraphics[width=0.95\linewidth]{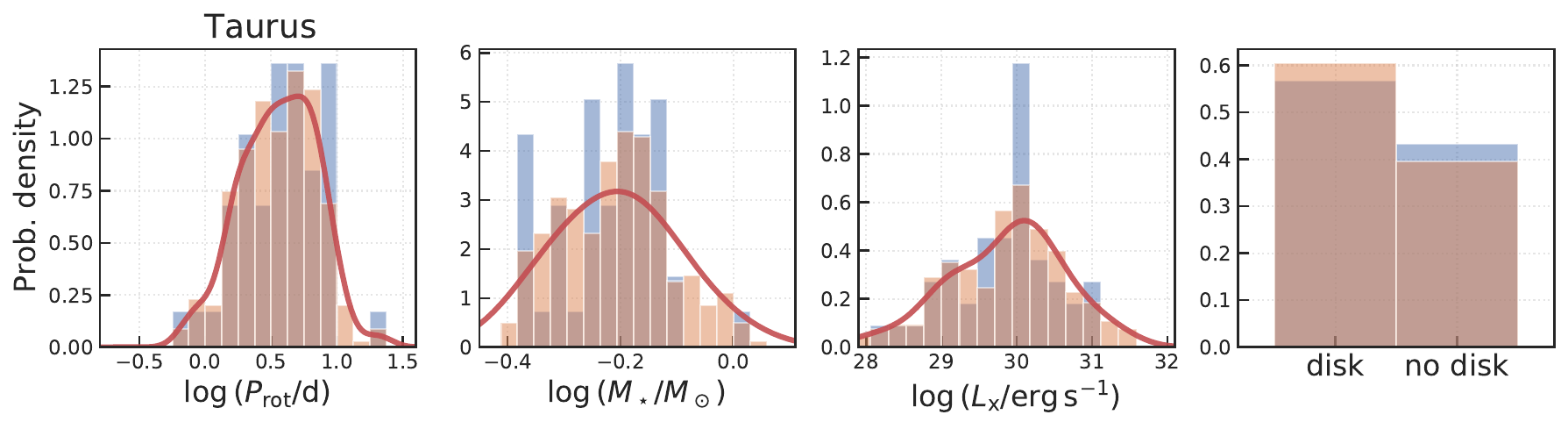}
    \caption{Calibration vs. resampled data sets used as input parameters for the rotation evolution model. The different panels show the logarithmic rotation period, stellar mass, X-ray luminosity and disk fraction. The blue histograms show the observed data, the red lines reflect the conditional probability density functions $P(\log\Prot)$, $P(\log \Mstar|\log \Prot)$, $P(\log \Lx|\log \Prot)$ and $P(f_\mathrm{disk}|\log \Prot)$, from which in total 300 data points were resampled, and whose distributions are shown as orange histograms. }
    \label{fig:resampled_properties}
\end{figure*}

As a next step we test the combined effect of the stellar mass and the disk-locking time using observed properties of star-forming regions as input parameters to our rotation evolution model. 
As previously described, our updated stellar spin-evolution model requires knowledge of the X-ray luminosity, $\Lx$, and the mass, $\Mstar$, of each individual star in order to calculate its corresponding inner disk lifetime. Further, an indicator for circumstellar gas or dust, $f_\mathrm{disk}$, (such as H$\alpha$- or NIR-excess emission) is required to assess if a given star is actually surrounded by a disk or not. Finally, for each of these stars a rotation period, $\Prot$, is needed as this is then used as the starting point from which the rotational evolution model is initiated.

For this purpose, we selected the 1--$3\,\Myr$ old Orion Nebula Cluster (ONC) and the Taurus molecular cloud complex, as they both belong to the best studied young star-forming regions close to the Sun, and additionally cover many extremes in environmental conditions. We obtained the final data set by cross-matching different observational catalogs, which is described in more detail in Appendix~\ref{appendix:catalog_compilation}.

However, these tight constraints on the observational data lead to a significant reduction of the final sample size compared to the individual stellar catalogs. We therefore create artificial samples using the observed data, which are used as calibration sets to calculate the conditional 2D kernel density estimates (KDEs) of the probability density functions $P(\log \Prot)$, $P(\log \Mstar|\log \Prot)$, $P(\log \Lx|\log \Prot)$ and $P(f_\mathrm{disk}|\log \Prot)$, from which in total 300 data points are then resampled. This greatly improves the sample size available to our study and thus the statistical significance of our results.
A comparison of the calibration and the resampled data sets is shown in Fig.~\ref{fig:resampled_properties}, where the blue histograms show the observed distributions, the red lines show the corresponding KDEs, and the orange histograms correspond to the resampled data sets. 
For completeness, the corresponding stellar mass (and X-ray luminosity) distributions of the resampled stellar parameters are shown and discussed in more detail in Appendix~\ref{appendix:resampled_properties}.

\begin{figure}[ht!]
    \centering    
    \includegraphics[width=0.95\linewidth]{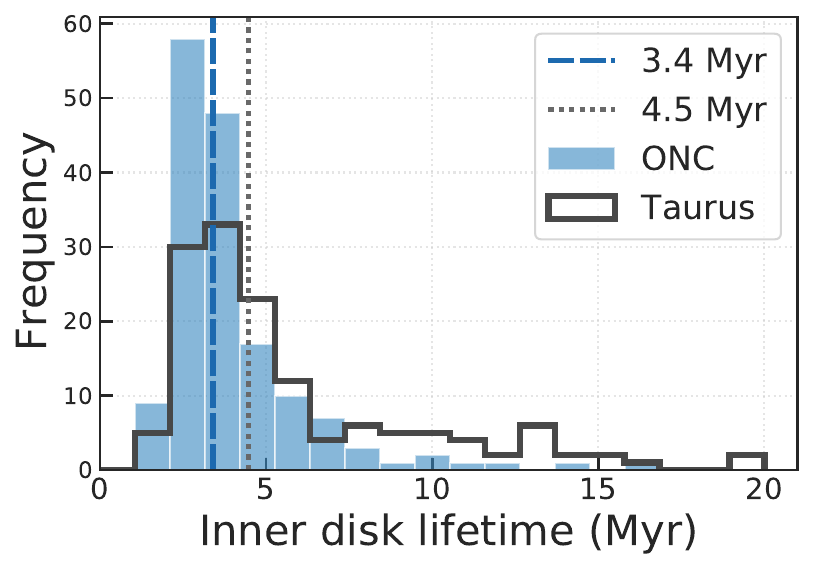}
    \caption{Distribution of calculated inner disk lifetimes for the ONC (blue) and Taurus (black). These were inferred based on the approach outlined in Sect.\,\ref{sec:methods_tdisc} using $\Mstar$ and $\Lx$ of each individual star. The dashed and dotted lines show the median inner disk lifetimes from the full stellar population of the ONC and Taurus investigated in our study.}
    \label{fig:disk_lifetimes_ONC_Taurus}
\end{figure}

Fig.~\ref{fig:disk_lifetimes_ONC_Taurus} shows the resulting distribution of disk-locking times that were obtained by following the approach outlined in Sect.~\ref{sec:methods_tdisc}. 
If a star has no circumstellar disk, which is assessed from the corresponding disk indicator shown in the last column of Fig.~\ref{fig:resampled_properties}, its rotational evolution is started at the mean age of the star-forming region, which in our study is assumed to be $t_0 = 2\,\Myr$. If a star has a disk, disk-locking is assumed and its rotation period is kept constant throughout the entire duration of the disk. Rotation period is then evolved self-consistently based on the rotational evolution model described in Sect.~\ref{sec:methods_rot}.

Due to the 2--3 order of magnitude scatter of $\Lx$ for given $\Mstar$ that is observed both in the ONC and Taurus \citep[cf.][]{Preibisch+2005, Guedel+2007}, the wide range of disk lifetimes that is also predicted from our model can be naturally recovered (see the lower panel in Fig.~\ref{fig:disc_life}). The inner disk lifetimes obtained reach values of up to $20\,\Myr$, with a more prominent tail towards longer disk lifetimes for Taurus due to its sub-population of stars with lower X-ray luminosities ($\log(\Lx/\ergs) \lesssim 28.5 $). This results in a $1\,\Myr$ longer median inner disk lifetime of $4.5\,\Myr$ for stars in Taurus compared to $3.4\,\Myr$ for stars in the ONC.

\begin{figure*}[t!]
    \centering    
    \includegraphics[width=0.49\linewidth]{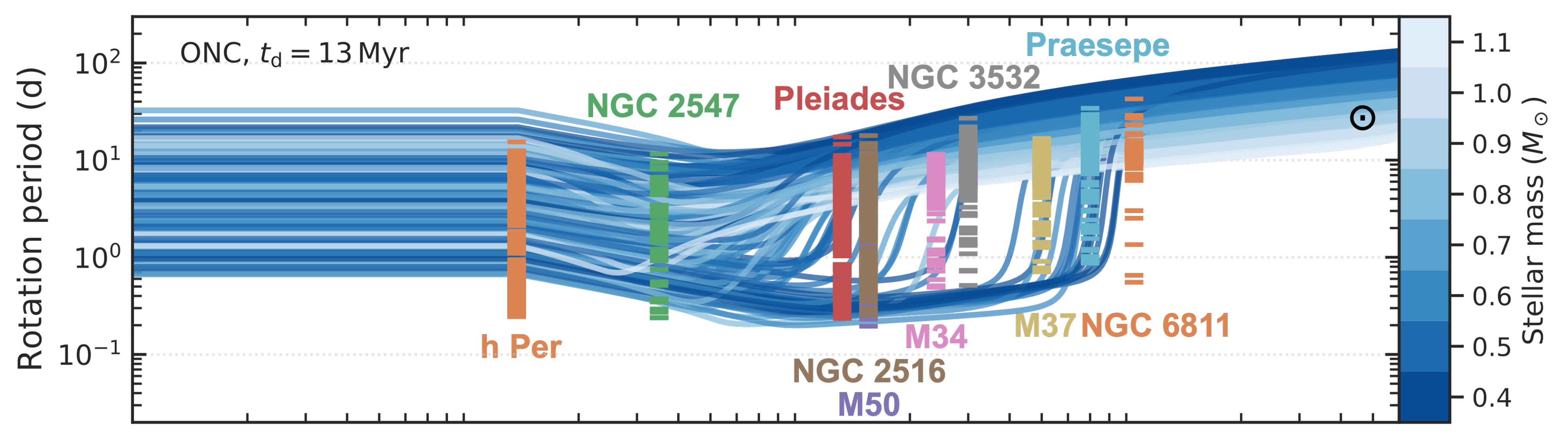}
    \includegraphics[width=0.49\linewidth]{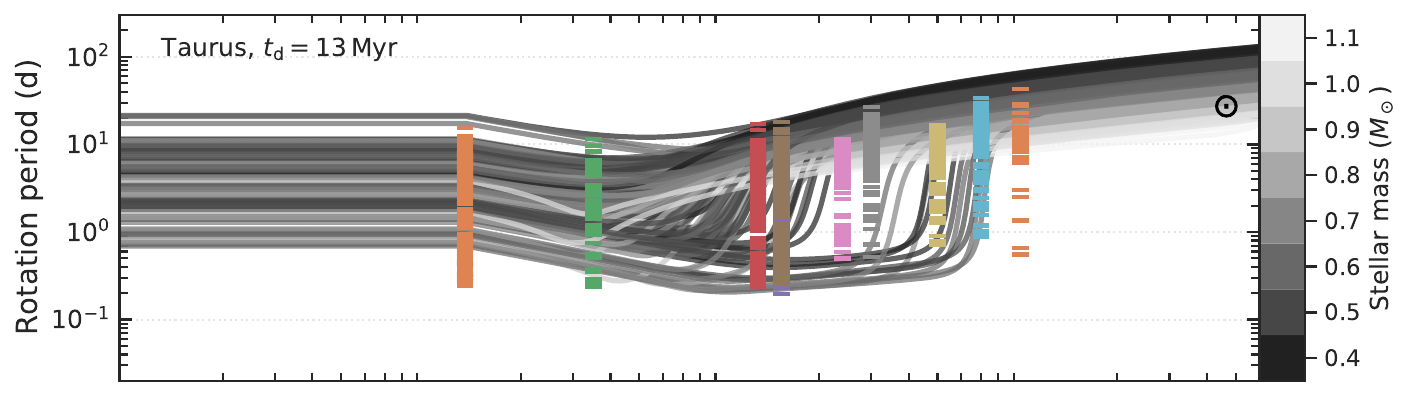}
    \includegraphics[width=0.49\linewidth]{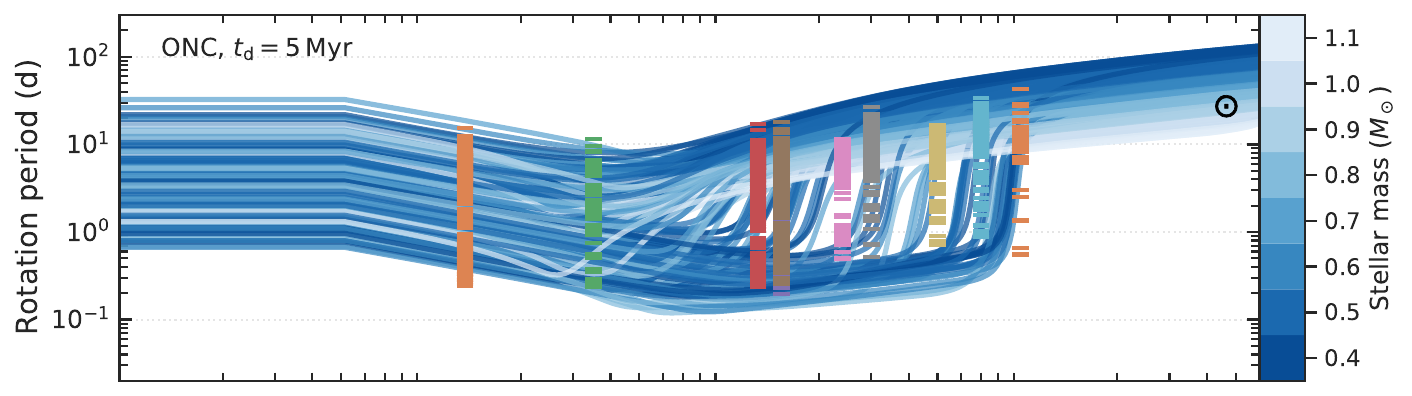}
    \includegraphics[width=0.49\linewidth]{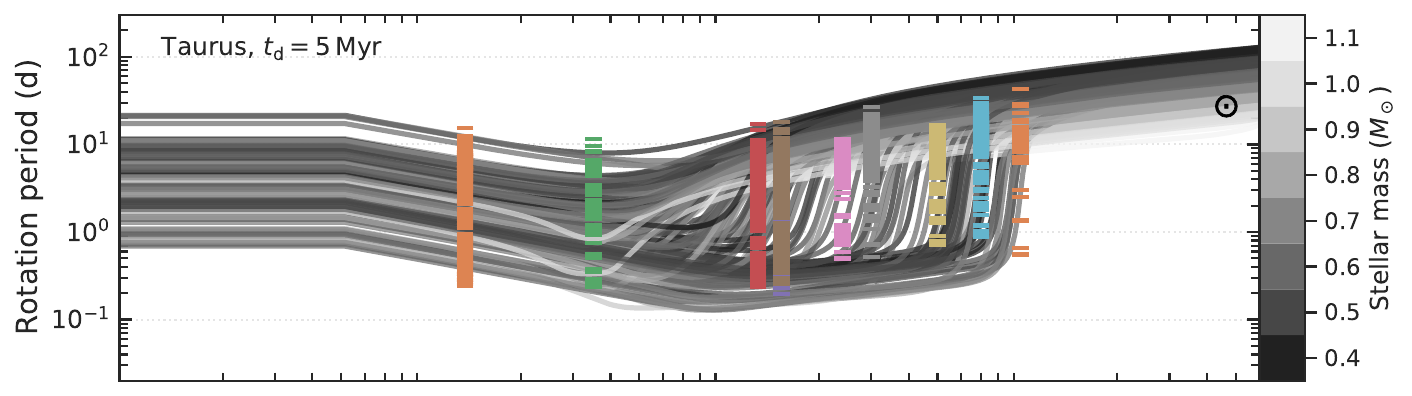}
    \includegraphics[width=0.49\linewidth]{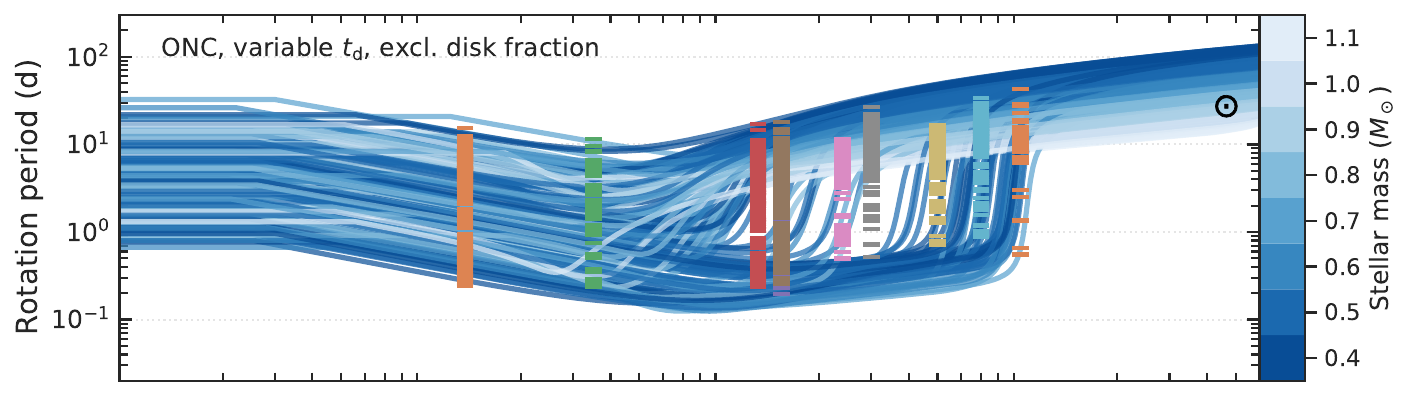}
    \includegraphics[width=0.49\linewidth]{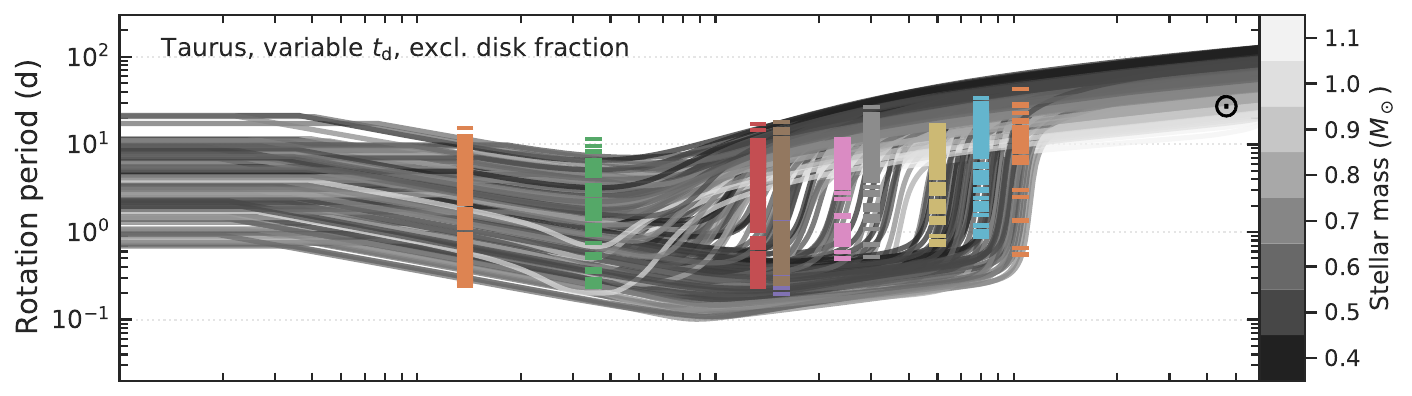}
    \includegraphics[width=0.49\linewidth]{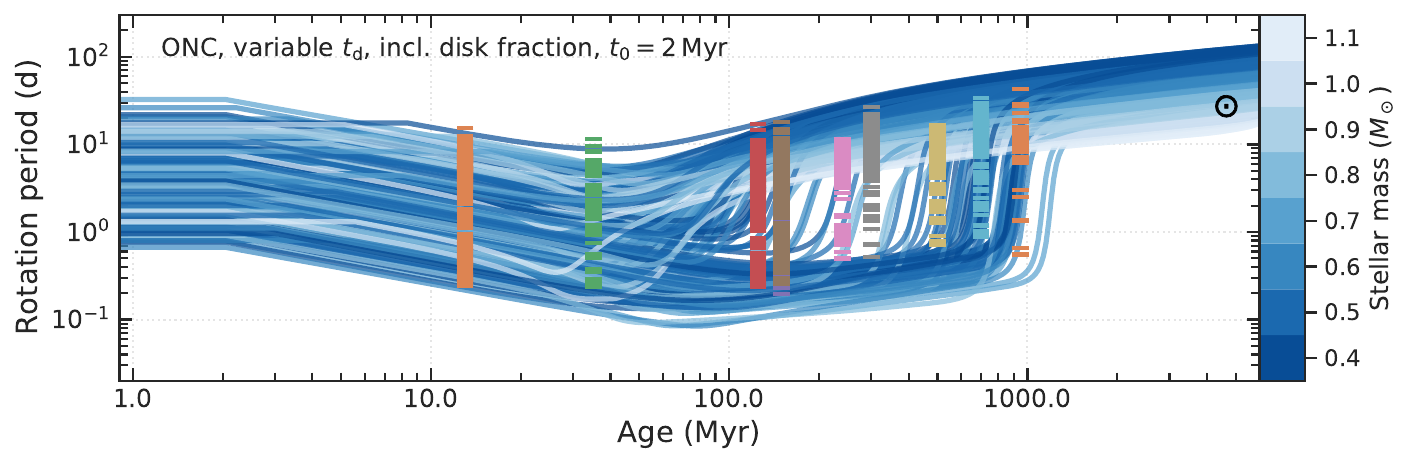}
    \includegraphics[width=0.49\linewidth]{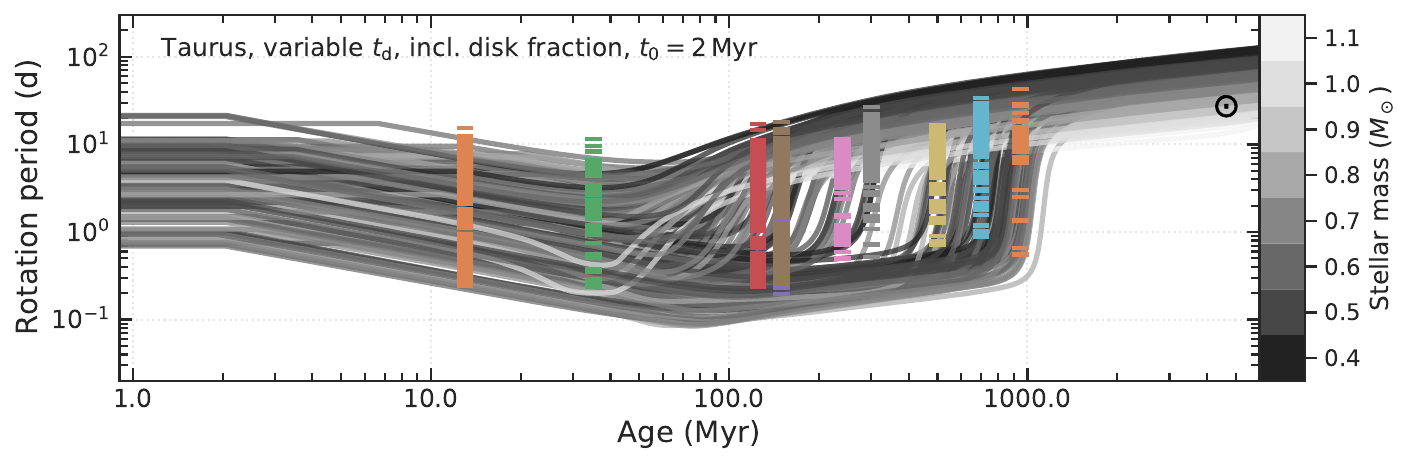}

    \caption{Time evolution of the stellar rotation periods assuming constant and variable inner disk lifetimes for the ONC (left) and Taurus (right), which are compared to open clusters of various ages.
    \textit{First row:} Model starting with a constant disk-locking time of $13\,\Myr$.
    \textit{Second row:} The same model assuming a shorter disk-locking time of $5\,\Myr$. 
    \textit{Third row:} The same model assuming variable disk-locking times that are calculated based on the stellar $\Lx$ and $\Mstar$. In this model, all stars are assumed to host a disk. 
    \textit{Fourth row:} The same as the previous model, but now the observed disk fractions, as shown in Fig.~\ref{fig:resampled_properties}, are taken into account. Disk-less stars are free to spin up immediately at $2\,\Myr$, which is the mean age of the ONC and Taurus that is assumed in our study, while the rotation of disk-bearing stars is braked and kept constant as long as they are surrounded by disks.}
    \label{fig:const_vs_var_SFRs}
\end{figure*}

Fig.~\ref{fig:const_vs_var_SFRs} shows the time-evolution of the stellar rotation periods, both for constant and variable inner disk lifetimes, using the ONC (left panels) and Taurus (right panels) as input distributions for our model. The color-coding of each evolutionary track reflects the stellar mass used, while the vertical, differently-colored distributions show the rotation periods of selected open clusters of different ages, which are listed in Table~\ref{tab:data_OC}.

In the case of constant disk-locking times, we chose $\tdisc=13\,\Myr$ and $5\,\Myr$ as two extreme cases. The former reflects the age of h\,Per, which is a young open cluster that is commonly used in similar studies as a starting point to rotational evolution models, as it is observed to have a low circumstellar disk fraction of $< 5\,\%$ \citep{Cloutier+2014}. 
The latter value of $5\,\Myr$ corresponds to a more typical mean disk lifetime that is inferred from a range of young star-forming regions (cf. Sec.~\ref{sec:intro_disklocking}) and is consistent with recent studies suggesting that the mean circumstellar disk lifetime can be significantly longer than the typically assumed 1--$3\,\Myr$ \citep[e.g.][]{Michel+2021, Pfalzner+2022}.

\begin{table}[t!]
    \centering
    \begin{tabular}{c|cl}
    \hline
        Name & Age & References\\
         & (Myr) & \\
        \hline
        ONC & 1-2 & \citet[][]{Getman+2005} \\
        & & \citet{Rodriguez-Ledesma+2009} \\
        & & \citet{Robberto+2020} \\
        Taurus & 1-3 & \citet[][]{Guedel+2007}\\
        & & \citet{Rebull+2020_Taurus} \\
        h\,Per & 13 & \citet{Moraux+2013}\\
        NGC\,2547 & $35$ & \citet[][]{GodoyRivera+2021} \\
        &  & \citet{Irwin+2008} \\
        Pleiades & $125$ & \citet[][]{GodoyRivera+2021} \\
        &  & \citet[][]{Rebull+2016a_Pleiades}\\
        &  & \citet[][]{Rebull+2016b_Pleiades}\\
        &  & \citet[][]{Stauffer+2016}\\
        M50 & $150$ &  \citet[][]{GodoyRivera+2021}\\
        NGC\,2516 & $150$ & \citet{Fritzewski+2020}\\
        & & \citet{TIC8.2}\\
        M34 & $240$ &  \citet{Meibom+2011}\\
        NGC\,3532 & $300$ & \citet{Fritzewski+2021}\\
        & & \citet{TIC8.2}\\
        M37 & $500$ & \citet[][]{GodoyRivera+2021}\\
        Praesepe & $700$ & \citet[][]{GodoyRivera+2021} \\
        & & \citet[][]{Douglas+2019}\\
        & & \citet[][]{Lodieu+2019}\\
        NGC\,6811 & $950$ & \citet[][]{GodoyRivera+2021} \\
        &  & \citet[][]{Curtis+2019}\\
        \hline
    \end{tabular}
    \caption{Open clusters with their adopted ages investigated in our study.}
    \label{tab:data_OC}
\end{table}

By construction, both the ONC- and Taurus-based models starting at a constant disk-locking time of $13\,\Myr$ fail at recovering the full distribution of rotation periods of h\,Per. This can be immediately understood, as these clusters have very different period distributions, with the ONC and Taurus showing on average slower rotation than h~Per. 
This implies that at the age of h\,Per, some significant dynamical evolution must have already taken place. 
Further, this model is also not able to reproduce the distributions of various other OCs, such as NGC~2547, NGC~2516 or NGC~6811. This confirms that mean disk lifetimes must be significantly shorter than $13\,\Myr$ in order to explain the wide distribution of rotation periods for OCs at $\sim 1\,\Gyr$.

Starting with a more realistic, shorter constant disk-locking time of $5\,\Myr$, as is shown in the second row of Fig.~\ref{fig:const_vs_var_SFRs}, a broader range of rotation periods for stars in h\,Per can be recovered, however, this model still falls short in reproducing its fastest rotators. 
Besides NGC~6811, the full range of OCs up to an age of $\sim 1\,\Gyr$ can now be recovered with this model.

The models assuming variable inner disk lifetimes are shown in the last two rows of Fig.~\ref{fig:const_vs_var_SFRs}, with the last one additionally taking circumstellar disk fractions of the corresponding star-forming region into account. 
Including variable inner disk lifetimes into the rotational spin-evolution model already significantly improves on the previous two cases with constant disk-locking times in terms of now fully being able to recover both ends of h\,Per's rotation period distribution while at the same now also matching the full range of rotation periods in NGC\,6811.

However, as was previously shown in Fig.~\ref{fig:resampled_properties}, only around 50--$60\,\%$ of the stars in our subsample of the ONC and Taurus are actually surrounded by circumstellar disks, meaning that at the current age of these regions, disk-less stars are expected to be already spinning up their rotation on their way towards the ZAMS. 
Therefore, the last model now also includes the observed disk fractions of the ONC and Taurus. Disk-less stars start evolving their rotation periods immediately at $t=2\,\Myr$, which is the mean age of the ONC and Taurus that we assume in this study, while the rotation of disk-bearing stars is kept constant until the full dispersal of their circumstellar disks.

Comparing the last two rows of Fig.~\ref{fig:const_vs_var_SFRs}, no significant difference in the overall evolution of the stellar population can be observed, confirming that it is indeed the inclusion of variable disk lifetimes times that lead to the improvement of the model in matching the full range of rotation periods of h\,Per, while simultaneously recovering rotation periods also for older open clusters, such as NGC~6811. 

However, Fig.~\ref{fig:const_vs_var_SFRs} also highlights a clear caveat of our model, which is the 
over-abundance of fast-rotating stars in older open clusters that it produces. One reason for this could be that mean disk lifetimes may be systematically longer than the $3$--$4\,\Myr$ obtained from our calculations \citep[e.g.][]{Michel+2021, Pfalzner+2022}. Fig.~\ref{fig:grid} shows that the number of stars in the fast-rotating branch increases with shorter disk lifetimes. Our calculation of disk lifetimes is based on the classical assumption that accretion onto the star is driven by viscous turbulence. If instead (weak) magnetic disk winds drive accretion, disk longevity may be increased significantly \citep[e.g.][]{Weder+2023}.

\begin{figure*}[t!]
    \centering    
    \includegraphics[width=\linewidth]{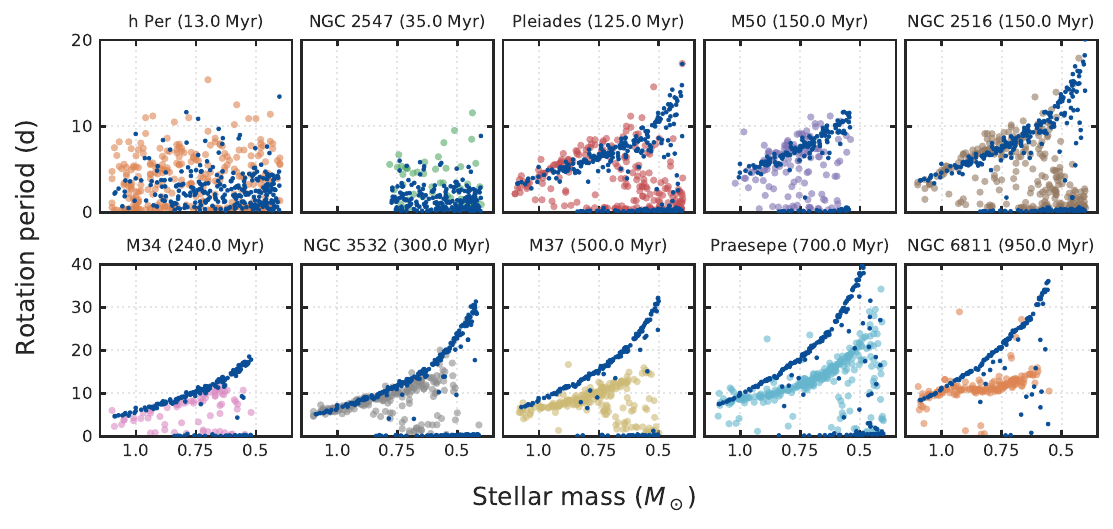}
    \includegraphics[width=\linewidth]{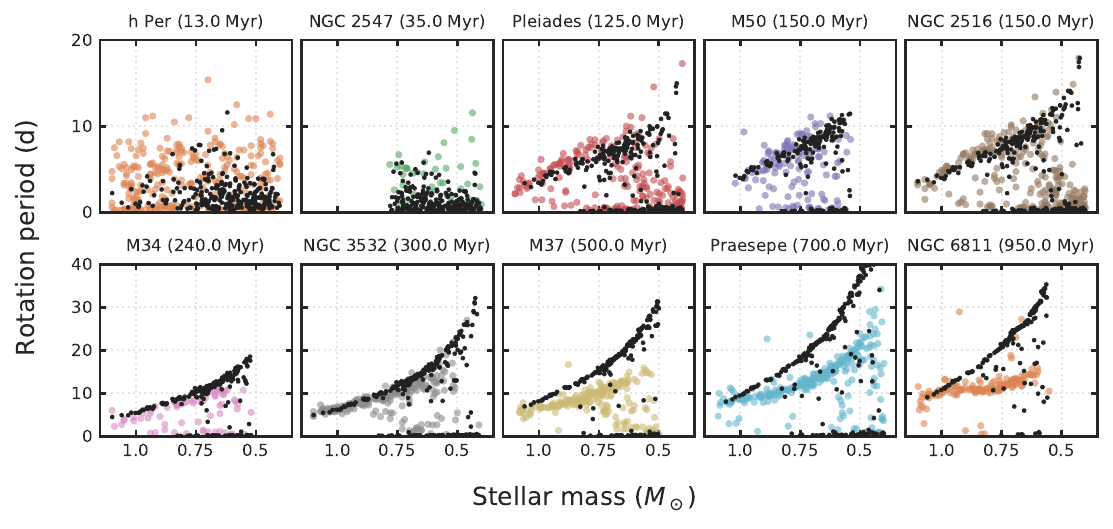}
    \caption{Comparison of stellar mass vs. rotation period distributions from our model evaluated at the ages of various open clusters. The model outputs shown here correspond to the variable disk lifetimes case including circumstellar disk fractions, shown in the last row of Fig.~\ref{fig:const_vs_var_SFRs}. Small, blue dots correspond to Orion, black dots to Taurus.} 
    \label{fig:mass_period}
\end{figure*}

While Fig.~\ref{fig:const_vs_var_SFRs} demonstrates our model's ability to cover the full range of rotation periods for various OCs of different ages, it is not useful to investigate if the aforementioned bimodality in rotation periods that is observed for many of these clusters, is also reproduced. 
Therefore, Fig.~\ref{fig:mass_period} now shows mass vs. rotation period diagrams of each OC shown in Fig.~\ref{fig:const_vs_var_SFRs} in direct comparison to the model distributions of the variable disk-locking time model including the disk fractions of the ONC and Taurus (i.e. the simulations shown in the last row of Fig.~\ref{fig:const_vs_var_SFRs}). At each age-cut, the mass range of our simulations is limited to the same as that of the corresponding OC to allow for a fair comparison.

In both cases, h~Per and NGC~2547 (note that the data presented here have no stars with $M_\star \gtrsim 0.75\,\Msun$) show more uniformly-distributed rotation periods across all masses that our model is able to emulate reasonably well (with the ONC-model performing better than the Taurus one, especially in the higher-mass regime). 
After $\gtrsim 100\,\Myr$, when most stars have reached the main sequence, a bimodality in rotation periods across all stellar masses begins to form. Stars with higher masses transition first from fast to slow rotation, thus recovering the mass dependence of the Skumanich branch. 
Also, the Skumanich branch becomes thinner with increasing age, due to more and more stars converging towards the $\Prot \propto t^{0.5}$ dependence. 

At about $500\,\Myr$, our model begins to predict systematically slower rotation than what is actually observed in OCs. This deviation of observed rotation periods from the standard Skumanich-relation in the form of stalled rotation implies that K- and M-dwarfs appear to spin down more slowly than F- and G-type stars \citep[e.g.][]{Agueros+2018, Curtis+2019, Curtis+2020}. A stellar mass-dependence of the rotational braking efficiency is to be expected \citep[see e.g.][]{Barnes2003, Barnes2007}, although this alone cannot explain the observed stalling of K-dwarfs with $\Prot \approx 20$--$40\,\mathrm{d}$, as is for example seen in the \textit{Kepler} field \citep[][]{McQuillan+2014}. 

Recent studies invoking a so-called core-envelope decoupling timescale between the star's radiative core and convective envelope that are assumed to rotate independently as solid bodies \citep[cf.][]{Spada+2011, SpadaLanzafame+2020} have been successful in reproducing the slow-rotator sequence, including the stalling of K- and G-dwarfs in regions like Pleiades, Praesepe, and NGC\,6811. The two-zone model can recover the overall trends in rotation observed in older open clusters, as well as \textit{Kepler} field stars, but to our knowledge has not been tested in order to simultaneously reproduce the bimodality including the fast rotator sequence in young open clusters, nor is it applicable to fully convective M-dwarfs.
This highlights the necessity of implementing more physical braking prescriptions into stellar rotational evolution models rather than evolving these solely based on empirical relations.

\subsection{Matching the rotation period distribution of h Per}
\label{sec:results_hPer} 

h\,Per, also known as NGC\,869, is a massive and young \citep[13--$14\,\Myr$,][]{Currie+2010} open cluster located in the constellation of Perseus and is part of the famous double cluster together with $\xi$ Persei (NGC\,884).
Due its fairly young age, which is just beyond the stage of circumstellar disk dispersal (10--$20\,\Myr$), it is an ideal laboratory for studying the rotational evolution of young stars over a wide range of stellar properties.
As stated previously, due to its low circumstellar disk fraction of 2--3\,\% with a slight increase to 4-5\,\% for subsolar-mass stars \citep[determined by measuring the $8\,\mu \mathrm{m}$ excess emission, cf.][]{Cloutier+2014}, h\,Per is often used as an input rotation period distribution for stellar spin-evolution studies.

\begin{figure*}[ht!]
    \centering    
    \includegraphics[width=1.\linewidth]{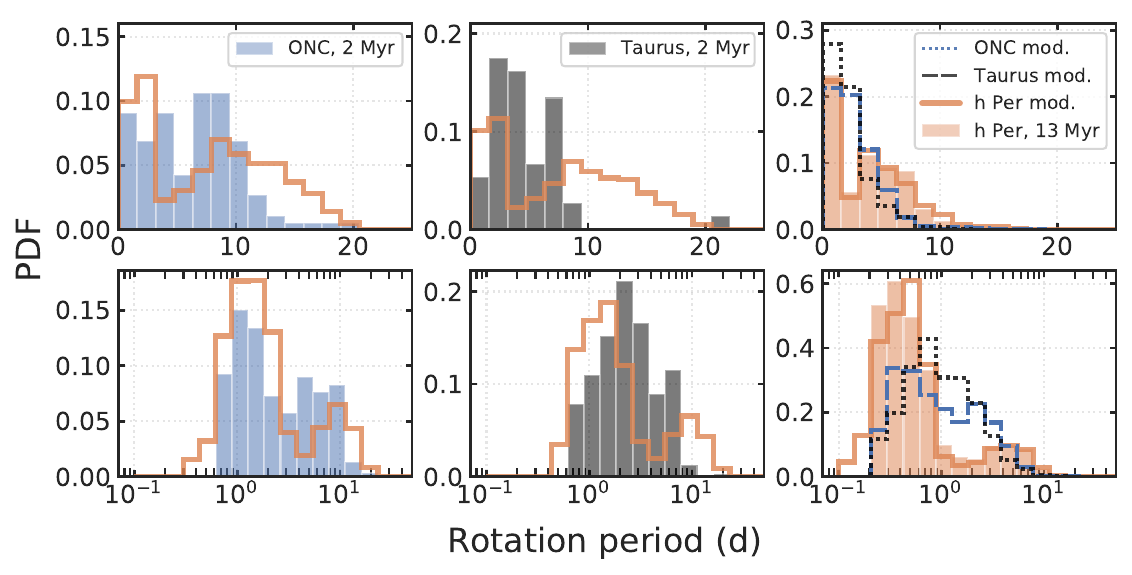}
    \caption{Comparison of observed rotation period distributions the ONC and Taurus with model outcomes assuming a synthetic initial rotation period in order to match h\,Per. The colored histograms show the $\Prot$ distributions of the ONC, Taurus and h\,Per in linear (top row) and logarithmic space (bottom row). The dashed blue and dotted gray lines in the right-hand column represent the models described in Sect.~\ref{sec:const_vs_var_SFRs} (last row of Fig.~\ref{fig:const_vs_var_SFRs}) for variable disk lifetimes, including the disk fraction for the ONC and Taurus, for which a mean age of $2\,\Myr$ was assumed. These models have been evolved to the age of h~Per. The solid orange line shows a synthetic $\Prot$-distribution with two Gaussian-shaped peaks centered on $2.2\pm1.5\,\mathrm{d}$ and $10\pm4\,\mathrm{d}$. The model tracks were each evaluated at the age of the corresponding region, i.e. at $2\,\Myr$ for the ONC and Taurus, and at $13\,\Myr$ for h\,Per. } 
    \label{fig:synthetic_Prot}
\end{figure*}

While we do find that including realistic disk-locking times improves the performance of our rotational evolution model in terms of matching the overall range of rotation periods for a number of different open clusters with various ages, at this point the model still falls short in reproducing the observed bimodality of rotation periods in h\,Per.
This is demonstrated in Fig.~\ref{fig:synthetic_Prot}, which shows a direct comparison between the observed distributions of the ONC, Taurus and h\,Per and the corresponding model distributions in linear (top row) and logarithmic space (bottom row). For this purpose, we selected the variable disk lifetime models for the ONC and Taurus (dashed blue and dotted gray lines, respectively) that include the disk fraction of the individual star-forming region, assuming a mean age of $2\,\Myr$ for both regions (i.e. row four in Fig.~\ref{fig:const_vs_var_SFRs}). 

When evaluated at the age of h\,Per (see last column in Fig.~\ref{fig:synthetic_Prot}), neither of the two models can successfully recover the exact distribution of rotation periods observed in h\,Per. 
Overall, the ONC model seems to perform better than the model using Taurus as starting point, as it can produce a bimodal rotation period distribution at an age of $13\,\Myr$, however the bimodality is much less pronounced than that of h\,Per, which appears to host a much larger fraction of fast-rotating stars.
On the other hand, the Taurus model results in an almost uni-modal distribution of rotation periods, which is an immediate result of the initial $\Prot$-distribution of Taurus being more uni-modal.

While the origin of the bimodality that is observed in h\,Per is not yet fully understood, similar studies to ours have made different attempts at understanding its population of fast-rotating stars \citep[e.g.][]{Vasconcelos+2015, Coker+2016, Roquette+2021, Vasconcelos+2022}.
Similar to \citet{Vasconcelos+2015}, we now produce a synthetic distribution of stellar properties in order to test how the unique shape of h\,Per's rotation period distribution can be recovered using our model. For this purpose, we create a double-peaked rotation period distribution, with two overlapping Gaussians centered on $2.2\pm1.5\,\mathrm{d}$ and $10\pm4\,\mathrm{d}$, respectively, which is shown as a solid orange line in Fig.~\ref{fig:synthetic_Prot} (`h\,Per mod.'). 
For each evolutionary track, the stellar mass is sampled from the initial mass function (IMF) derived from the ONC \citep[][]{DaRio+2012}, and the corresponding X-ray luminosity is subsequently derived from the stellar mass using the XLF observed in the ONC \citep[][]{Preibisch+2005}. 
Additionally, we need to invoke an initial disk fraction of $65\,\%$, with the stars in the fast-rotating peak being initially defined as `disk-less', so that they immediately start spinning up at $t_0=2\,\Myr$. The slow-rotating stars are defined as `disk-bearing', and are thus braked for the entire duration of their corresponding disk lifetime. Using this synthetic distribution, the overall characteristics of h\,Per can now be recovered almost perfectly.

The synthetic $\Prot$-distribution shares some broad similarities with the one observed for ONC, for example that it is double peaked. For the ONC, however, the peaks are observed to lay at $2\,\mathrm{d}$ and $8\,\mathrm{d}$ \citep[][]{Herbst+2002} and are thus located closer to each other. Further, the synthetic distribution extends farther towards both smaller and higher rotation periods. 
In contrast, the distribution observed in Taurus appears to be very different to the synthetic one, suggesting that h\,Per and Taurus stem from two very different sets of initial conditions that have shaped their rotation period distributions, perhaps pointing to different modes of star formation. Instead, the ONC and h\,Per likely share more similarities in their early evolution.

The fact that at an age of only $2\,\Myr$ the ONC already hosts a bimodal distribution in rotation periods suggests that some significant disk evolution must have taken place. It
is known that very young stellar clusters close to the Class~0/I phase , such as the $\sim 0.3\,\Myr$ old NGC\,2024 stellar cluster \citep[which has a disk fraction of $86\pm 8\,\%$, cf.][]{Haisch+2000} can already show a significant reduction in their disk fraction. Such clusters are generally too young for the disks to have been destroyed by planet formation and/or internal photoevaporation only, but the reduced disk fraction can be attributed to binary interactions. 
Nevertheless, an initial disk fraction of only $65\,\%$ at an age of 1--$2\,\Myr$ for a young star-forming region can only be achieved if a subset of stars has been subject to an additional disk-destroying mechanism, such as external photoevaporation due to the irradiation from nearby OB-stars. 
This is consistent with the recent study by \citet{Roquette+2021}, who find that the excess of fast rotators in h\,Per can be explained by a high FUV environment during the cluster's early PMS evolution.

Consequently, while in this study, we solely focused on internal photoevaporation due to coronal EUV/X-ray emission, the combined effect of internal \textit{and} external photoevaporation onto the disk dispersal process as well as the subsequent rotational evolution of their host stars should be investigated in more detail in the future.

\section{Model limitations} 
\label{sec:discussion}

\subsection{External photoevaporation in Orion and Taurus}
\label{sec:external_PE}

We have purposely ignored the effect of external photoevaporation in our study, even though it has been found to play an important role in the evolution of at least some protoplanetary disks, which are embedded in clusters with high UV-radiation fields. 

The ONC corresponds to an immensely dense star-forming environment, with a central stellar density of $\sim 4700\,\pc^{-3}$ \citep[][]{King+2012}, and is known to host a strong FUV radiation environment \citep[cf.][for a review]{WinterHaworth2022}. This is likely to affect the star- and planet-formation process through the external photoevaporation of circumstellar disks by the massive OB-stars hosted by the ONC.
In contrast, Taurus' total mass and stellar number density are significantly smaller \citep[$n\sim 1$--$10\,\pc^{-3}$, e.g.][]{Briceno+2002, King+2012}, meaning that Taurus corresponds to the prototypical test bed for studying non-clustered star formation, in which stars are forming in relative isolation, free from the vigorous UV irradiation coming from nearby massive stars, thus explaining the more unimodal shape of its rotation period distribution.

While strong UV radiation fields may significantly truncate outer disk radii, as has been observed in Orion (the so-called proplyds), their effect on the inner disk (i.e. inside of the gravitational radius), which determines the amount of gas that is ultimately accreted onto the central star and thus determines the inner disk lifetime, remains unclear. 
If the star-forming environment was indeed to be significantly affecting the disk dispersal process, one would expect to observe systematically smaller disk fractions in dense stellar clusters with strong UV-radiation fields compared to less-densely populated clusters of the same age. Observed disk fractions of young star-forming regions may hint towards such a correlation \citep[e.g.][]{Ribas+2014}, however, drawing a definitive conclusion is difficult, especially due to the large uncertainty in their ages in the order of a factor 2 \citep[e.g.][]{Bell+2013}. 

Stars are expected to be born with a range of possible rotation velocities, which can probably be related to different environmental factors (e.g. location within the molecular cloud, abundance of nearby massive stars, stellar encounters, etc.).
In a recent study by \citet{XiaoChang2018}, it was proposed that prestellar cores with initially small rotational velocities seem to be less affected by external photoevaporation (even in high FUV-radiation fields), as most of the mass will be accreted onto the star rather than being deposited in the disk's outer region. Fast rotating prestellar cores on the other hand seem to be strongly affected by external photoevaporation. 

However, the FUV radiation field strength strongly decreases as a function of distance from the massive OB-stars. This is confirmed by recent (sub-)mm observations of the Orion~A molecular cloud, that find circumstellar disk masses to be very similar to those of nearby low-mass star-forming regions at similar ages \citep[e.g.][]{vanTerwisga+2022}. This means that only disks located very close to the Trapezium cluster are expected to be significantly affected by the highly-energetic radiation emitted by the OB-stars. 

In this context, \citet{Richert+2018} assembled a large unified data set that gathers ages, stellar masses, and disk fractions of more than 69 young stellar clusters. They did not find clear evidence that disk longevity indeed depends on the surrounding star-forming environment. 
Thus, more work on the influence of environmental conditions on the disk dispersal process that are combined with realistic prescription of internal photoevaporation as well as MHD winds is urgently needed.

\subsection{Constant rotation and X-ray activity during disk-locking}

While our model is successful in reproducing the rotational evolution of a population of stars qualitatively, it is still suffering from many simplifications that we have assumed in our study.

While we improve on previous by considering a non-uniform duration of the disk-locking phase, we nevertheless assume that during this time, the stellar rotation rate stays constant, i.e. the star is `locked' to its surrounding disk, as is suggested by observations of young star-forming regions. The study by \citet{Gallet+2019} tested if this is indeed a realistic assumption. They find that the stellar rotation rate may possibly increase during the disk locking time and in order for the rotation rate to stay constant, high magnetic field strengths of a few Kilogauss would be required.

Similarly, we assume constant X-ray luminosities throughout the PMS phase. In contrast, \citet{Preibisch+2005} found a mild decay of the X-ray luminosity during the first $10\,\Myr$, scaling as $\Lx \propto \tau^{-1/3}$, whereas \citet{Getman+2022} find a decay of $\Lx \propto \tau^{-1/2}$ between 7--$25\,\Myr$ for solar-type stars.
While this is expected to only have a minor impact on our model, this decrease of $\Lx$ would lead to longer disk lifetimes, and thus the stars would be disk-braked, on average, for a longer duration. 
As could be inferred from Fig.~\ref{fig:grid}, longer disk lifetimes for a given stellar mass would mainly impact the rotational evolution between $100\,\Myr$ to $1\,\Gyr$, in which the bimodality in $\Prot$ is observed. Therefore, our conclusion would qualitatively remain the same, even though the evolution of individual stars may be different.
\section{Summary and conclusions} 
\label{sec:conclusions}

In this paper, we have explored the impact of circumstellar disk dispersal via EUV/X-ray-photoevaporation on the rotational evolution of low-mass stars. In particular, we focused on assessing how the implementation of a distribution of realistic inner disk lifetimes is influencing the rotation period distribution of a population of stars with varying input parameters. The main results can be summarized as follows:

\begin{enumerate}
    \item Employing a state-of-the-art EUV/X-ray disk photoevaporation model we have derived a simple recipe for the calculation of circumstellar disk lifetimes inside of their gravitational radius that is solely dependent on the stellar mass and the X-ray luminosity during the PMS phase. Using stellar properties from the Orion Nebula Cluster and the Taurus star-forming region, we obtain mean disk-locking times of $3.4\,\Myr$ and $4.5\,\Myr$, respectively. 
    
    \item This prescription for calculating realistic disk-locking times was then implemented into a magnetic-morphology-driven rotational evolution model. We find that the duration of disk-locking can have a strong impact on the subsequent rotational evolution of a given star by counteracting its spin-up due to contraction during the PMS. 
    Initially slow-rotating stars that are braked by their disks for a longer period of time tend to remain slow rotators for longer compared to coeval stars that are braked for a shorter duration of time. 

    \item By assuming a realistic distribution of inner circumstellar disk lifetimes for stars in the ONC or Taurus, the range of rotation periods for older clusters, such as Pleiades, Praesepe or NGC~6811 can be easily recovered in contrast to the models that assume constant inner disk lifetimes. 
    
    \item For all young star-forming regions that were tested with our model, the strong bimodality in rotation periods that is observed in h\,Per cannot be recovered, however. 
    By producing a synthetic rotation periods distributions with two Gaussian peaks located at $2.2\pm1.5\,\mathrm{d}$ and $10\pm4\,\mathrm{d}$, as well as assuming an initial disk fraction of $65\,\%$, we can successfully reproduce h\,Per. At an age of $\sim 1$--$2\,\Myr$ for the ONC and Taurus, such a low disk fraction can only be achieved if an additional disk dispersal process, such as external photoevaporation, is invoked. 
    Future work should therefore focus on investigating the combined effect of both internal and external photoevaporation on the rotational evolution of young stars. 
    
\end{enumerate}

\begin{acknowledgments}
We thank the referee for an insightful review, which helped to improve the clarity of this paper.
KM was supported by NASA {\it Chandra} grants GO8-19015X, TM9-20001X, GO7-18017X, and HST-GO-15326.
JJD and CG were funded by NASA contract NAS8-03060 to the {\it Chandra X-ray Center} and thank the Director, Pat Slane, for continuing advice and support. BE and GP acknowledge the support of the Deutsche Forschungsgemeinschaft (DFG, German Research Foundation) - 325594231. This research was supported by the Excellence Cluster ORIGINS which is funded by the Deutsche Forschungsgemeinschaft (DFG, German Research Foundation) under Germany´s Excellence Strategy – EXC-2094-390783311. 
\end{acknowledgments}


\software{\texttt{astropy} \citep{astropy:2013, astropy:2018, astropy:2022},  
          \texttt{matplotlib} \citep{Hunter2007},
          \texttt{numpy} \citep{numpy},
          \texttt{read\_mist\_models} \citep{Choi+2016, Dotter+2016},
          \texttt{scipy} \citep{scipy},
          \texttt{seaborn} \citep{seaborn}
          }

\appendix
\section{Calculation of disk lifetimes}
\label{sec:appendix_tdisc}

The inner disk lifetime, $\tdisc$, can be determined by calculating the so-called `clearing timescale', $\tclear$, which corresponds to the time at which photoevaporation starts to clear the circumstellar disk, i.e. where $\Mdotwind (\Mstar, \Lx) \approx \Mdotacc(t) $ \citep{Clarke+2001, Ruden2004}. 
To this value we add the viscous timescale $\tnu$ at the gravitational radius $R_\mathrm{g} \approx 8.9\,\au ( \Mstar/\Msun )$, which describes the location from which photoevaporative winds can be established \citep{Owen+2012}, so that:

\begin{equation}
    \tdisc = \tclear (R_1) + \tnu (R_\mathrm{g}), 
\end{equation}
where
\begin{equation}
    \tclear = \frac{t_\mathrm{\nu}(R_1)}{3} \left( \frac{3 M_\mathrm{d,0}}{2t_\mathrm{\nu}(R_1) \dot{M}_\mathrm{w}(\Lx)} \right)^{2/3},
\end{equation}
and 
\begin{equation}
    \tnu = \frac{R^2}{3\nu}.
\end{equation}
Here, $R_1$ corresponds to the exponential cut-off radius that is typically located somewhere between 50--$100\,\au$ \citep[cf. Eq. 3 in][]{Clarke+2001}, $M_\mathrm{d,0}$ is the initial disk mass, $\dot{M}_\mathrm{w}$ is the mass-loss rate due to photoevaporative winds and $\nu$ is the disk's viscosity.  

We subsequently follow \citet{Picogna+2021} and adopt the $\Mdotacc$--$\Mstar$-relationship for the 1--$3\,\Myr$ old Lupus star-forming region \citep[][]{Alcala+2017, Alcala+2021} to obtain observationally motivated stellar accretion rates:

\begin{equation}
    \log (\Mdotacc/\Msun\,\yr^{-1}) = 
  \begin{cases}
    4.58(\pm0.68) \log \left(\Mstar/\Msun\right) - 6.11(\mp0.61) & \text{if  } \Mstar \leq 0.2\Msun\\
    1.37(\pm0.24) \log \left(\Mstar/\Msun\right) - 8.46(\mp0.11) & \text{otherwise}.
  \end{cases}
\end{equation}
From these, the initial viscous accretion rates $\dot{M}_\mathrm{acc,0}$ can be derived by assuming a power-law scaling of the accretion rate with time \citep[e.g.][]{Lynden-BellPringle1974, Hartmann+1998}: 
\begin{equation}
    \Mdotacc(t) \approx \dot{M}_\mathrm{acc,0} \left( \frac{t}{\tnu} \right)^{-3/2} = \frac{M_\mathrm{d,0}}{2t_\nu} \left( \frac{t}{\tnu} \right)^{-3/2},
\end{equation}
and from which follows that $\tnu=M_\mathrm{d,0}/(2\dot{M}_\mathrm{acc,0})$. The initial disk mass is estimated from the stellar mass via $M_\mathrm{d,0}=0.14\,\Mstar$, which corresponds to the best fit to observed disk lifetimes in the $\sim 5\,\Myr$ old $\lambda$Ori open cluster \citep[][]{Picogna+2019}.

The mass loss rate, $\Mdotwind$, of a circumstellar disk around a star with $\Mstar = 0.7\,\Msun$ as a function of X-ray luminosity in the `soft' part of the X-ray spectrum, $L_\mathrm{x,soft}$, is described by: 
\begin{equation}
\label{eq:Mdotwind_soft}
    \log \Mdotwind = a_\mathrm{L}  \exp{\left(\frac{ [\ln{(\log{L_\mathrm{X, soft}})} - b_\mathrm{L}]^2}{c_\mathrm{L}}\right)} + d_\mathrm{L},
\end{equation}
where $a_\mathrm{L}=-1.947\times10^{17}$, $b_\mathrm{L}=-1.572\times10^{-4}$, $c_\mathrm{L}=-2.866\times10^{-1}$ and $d_\mathrm{L}=-6.694$ \citep[cf. Eq.~9 in][]{Ercolano+2021}.\footnote{We use the notation $\ln$ for the natural logarithm with base $e$, and $\log$ for the decimal logarithm with base 10.}
By obtaining a linear fit between the nominal and soft X-ray luminosities given in Table~4 of \citet{Ercolano+2021}, $L_\mathrm{X, soft}$ can then be determined from the total X-ray luminosity, $\Lx$, via the linear relation:
\begin{equation}
    \log L_\mathrm{X, soft} = 0.95 \,\log \Lx + 1.19.
    \label{eq:Lxsoft}
\end{equation}

While these equations take the dependence of $\Mdotwind$ on the stellar X-ray luminosity into account, they do not consider the dependence of the X-ray luminosity on the stellar mass. \citet{Owen+2012} found $\Mdotwind$ to be almost independent of the stellar mass (with $\Mdotwind \propto \Mstar^{-0.068}$), however the more recent study by \citet[][see their Eq.~5]{Picogna+2021} finds a linear dependence:
\begin{equation}
\label{eq:Mdotwind_Mstar_variableLx}
    \Mdotwind (\Mstar) = 3.93\times 10^{-8} \left( \frac{\Mstar}{\Msun}\right),
\end{equation}
which is based on radiation-hydrodynamical simulations that assumed the \textit{median} X-ray luminosity for each modelled stellar mass, derived from the \textit{XMM-Newton} Extended Survey of the Taurus Molecular Cloud \citep[XEST,][]{Guedel+2007}: 

\begin{equation}
\label{eq:Guedel}
    \log L_\mathrm{X,med}=(1.54 \pm 0.12) \log \left( \frac{\Mstar}{\Msun}\right) + (30.31 \pm 0.06).
\end{equation}
The mass loss rate as a function of the soft X-ray luminosity \textit{and} stellar mass can then be calculated via:

\begin{equation}
\label{eq:Mdotwind_Mstar_Lx}
    \Mdotwind (\Mstar, L_\mathrm{X,soft}) = \Mdotwind (\Mstar) \frac{\Mdotwind(L_\mathrm{X,soft})}{\Mdotwind(L_\mathrm{X,soft,med})},
\end{equation}
which is simply Eq.~\ref{eq:Mdotwind_Mstar_variableLx} multiplied by Eq.~\ref{eq:Mdotwind_soft}, normalized by the wind mass loss rate for the \textit{median} soft X-ray luminosity as well as by exploiting Eq.~\ref{eq:Lxsoft} \citep[see also][for a similar approach]{Burn+2022}.

Using this approach, the inner disk lifetime of a given star can be estimated by solely providing its observed X-ray luminosity $\Lx$ and stellar mass $\Mstar$ as an input, while all remaining initial disk parameters can be derived from observationally motivated scaling relations:

\begin{equation}
    \tclear = \frac{M_\mathrm{d,0}}{2\dot{M}_\mathrm{acc,0}^{1/3} \dot{M}_\mathrm{w}^{2/3}}, 
\end{equation}
\begin{equation}
    \tnu = \frac{M_\mathrm{d,0}}{2\dot{M}_\mathrm{acc,0}}.
\end{equation}

\section{Compilation of observational data}
\label{appendix:catalog_compilation}

\subsection{Orion Nebula Cluster (ONC)}
\label{sec:ONC}

The Orion Nebula Cluster (ONC) is a young \citep[1-2 Myr,][]{Fang+2021} and massive star-forming region located in the Orion Molecular Cloud Complex. With a distance of $414 \pm 7\,$pc \citep[][]{Menten+2007}, the ONC resembles the closest massive star-forming region and is thus a prime target for studies of PMS stars and their evolution. 

We use X-ray data from the \textit{Chandra} Orion Ultradeep Project \citep[COUP, cf.][]{Getman+2005, Preibisch+2005}, that performed the deepest and longest X-ray observation of the ONC so far. The COUP catalog comprises more than 1616 individual sources, with most of them having optical and near-infrared counterparts.
The COUP data were then cross-matched with the catalogs by \citet{Davies+2014} and \citet[][]{CiezaBaliber2007}, who collected measured rotation periods as well as circumstellar disk indicators of PMS stars in the ONC from various sources in the literature.\footnote{We note, that while \citet{Serna+2021} obtained the most recent collection of rotation properties for young PMS stars within the Orion Star-Forming Complex (including regions such as the ONC, $\sigma$\,Ori, $\lambda$\,Ori and 25\,Ori), we refrain from using their sample. Their study only provides $v\sin i$ measurements that were determined using TESS light curves from which it is intrinsically difficult to determine precise deprojected rotation periods in clustered young stellar regions using this method (which are, however, needed as an input for our model).} 
Finally, the resulting data were additionally cross-matched with the catalog published by \citet{Robberto+2020}, who determined updated stellar masses for the PMS population in the ONC.

\subsubsection{Taurus Molecular Cloud}
\label{sec:Taurus}

The Taurus-Auriga star-forming region (or simply Taurus) is a nearby low-mass star-forming region located about $141\pm7\,$pc away from Earth \citep[][]{Zucker+2019} and is similarly young as the ONC with an estimated age of about $\sim2\,\Myr$.

We use X-ray data from the \textit{XMM-Newton} Extended Survey of the Taurus molecular cloud \citep[XEST,][]{Guedel+2007}, which collected deep X-ray observations of Taurus' most populated $\sim 5\,$deg$^2$ (covering an area that is about 60 times larger than the one observed in the COUP-survey).
These were then cross-matched with photometric rotation periods measured by \citet[][]{Rebull+2020_Taurus} using light curves from the \textit{K2} mission, as well as \citet[][]{Davies+2014}, who both list the (non-)presence of a circumstellar disk. 
In order to have a stellar mass for every star in our sample, we additionally cross-matched the resulting catalog with the TESS input catalog v8.2 \citep[TIC, ][]{TIC8.2}, which is a compiled catalog of stellar parameters for every optically persistent, stationary object in the sky.

\subsubsection{Final observational sample}

The resulting samples of the ONC and Taurus are not necessarily representative of their actual stellar population as the data points were selected based on having a measured rotation period, stellar mass, X-ray luminosity \textit{and} circumstellar disk indicator. These parameters are all derived from different observational surveys, which all have their own selection effects and biases. 
However, as the ONC and Taurus belong to the best studied star-forming regions (both having been studied deeply in various X-ray, IR and mm campaigns), we do not expect these selection effects to significantly impact our inferences.

\section{Mass distribution of resampled stellar properties}
\label{appendix:resampled_properties}

\begin{figure}
    \centering
    \includegraphics[width=\linewidth]{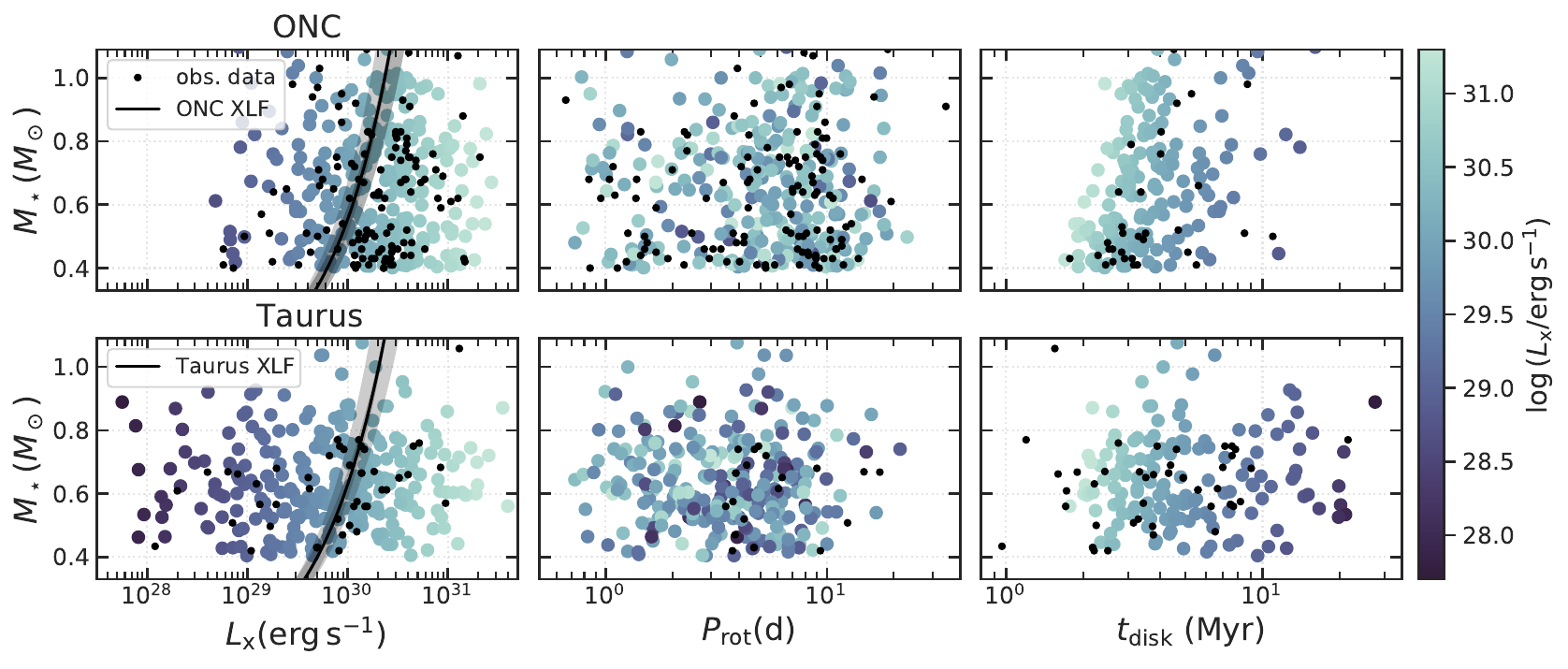}
    \caption{Stellar mass distributions of the observed data (black dots), as well as the resampled distributions for the ONC (top row), and Taurus (bottom row), that were used for the here presented rotational evolution model. For the resampled data sets, the color-coding reflects the X-ray luminosity of each star. The first column additionally shows the X-ray luminosity functions for the ONC \citep{Preibisch+2005} and Taurus (see Eq.~\ref{eq:Guedel}) with their respective uncertainties as black lines.}
    \label{fig:mass_distribution}
\end{figure}

In Fig.~\ref{fig:resampled_properties} we have shown the individual distributions of the resampled stellar properties that were used as an input to our rotational evolution model. Fig.~\ref{fig:mass_distribution} now shows the mass distribution of these parameters the ONC (top row) and Taurus (bottom row), where the color-coding reflects the X-ray luminosity of each individual star. In addition, we have overplotted the original data for both regions that were used as calibration sets as black dots.

The first column shows the stellar mass vs. X-ray luminosity distribution.
A strong scatter spanning several orders of magnitude in X-ray luminosity for given stellar mass can be observed, which is in agreement with previous studies of X-ray properties of young star-forming regions. However, there  does not appear to be strong correlation of $\Lx$ with $\Mstar$, which may be unexpected given the clear trends inferred from XLFs of e.g. ONC \citep{Preibisch+2005} or Taurus (Eq.~\ref{eq:Guedel} and \citealt{Guedel+2007}) that are overplotted in the first column with their corresponding uncertainties as shaded regions. 
This discrepancy can be easily explained as the stellar samples considered in our study are not entirely reflective of the actual stellar populations. The reason for this is simple--we required all stars in our sample to have a measurement of the stellar mass, X-ray luminosity, rotation period \textit{and} a disk presence indicator, while the samples studied e.g. by \citet{Preibisch+2005} or \citet{Guedel+2007} only required measurements of the X-ray luminosity and stellar mass in order to be able to infer an XLF.
Further, no correlation of rotation periods with stellar mass nor X-ray luminosity can be observed in this data set. 

Only a clear trend towards longer disk lifetimes for higher-mass stars is observed that, however, appears to be in contrast with the previously reported result of longer mean disk lifetimes for lower-mass stars (see the lower panel of Fig.~\ref{fig:disc_life}). This apparent discrepancy can be readily explained by the fact that 1) lower mass stars have---on average---lower-mass disks and 2) that these systems with the shortest disk lifetimes consistently correspond to stars with the highest X-ray luminosities. A low-mass disk is much more susceptible to internal photoevaporation than a high-mass disk for given $\Lx$, which therefore results in shorter lifetimes for lower-mass stars. 
In contrast, Fig.~\ref{fig:disc_life} shows that the \textit{median} disk lifetimes increase systematically for lower-mass stars when their \textit{mean} X-ray luminosities are used to compute the disk lifetimes.

Another clear trend that can be observed in Fig.~\ref{fig:mass_distribution} is the dependence of the disk lifetimes on the stellar X-ray luminosity. This trend is expected from our model as $\Lx$ is the dominant factor in determining the photoevaporation rate for a given star and thus the efficiency of XPE.
The lower $\Lx$, the longer therefore the disk lifetime, ultimately resulting in disk lifetimes of more than 10--$20\,\Myr$ for stars where $\log(\Lx/\ergs) \lesssim 29$. Due to the previously discussed scatter of $\Mstar$ with $\Lx$, such long disk lifetimes can therefore be achieved for all stellar masses meaning that it is not surprising to observe stars with such long disk lifetimes in our simulated samples of the ONC or Taurus.

\bibliography{literature}{}
\bibliographystyle{aasjournal}



\end{document}